\documentclass[%
 reprint,
 superscriptaddress,
 numerical,
 showpacs,
 amsmath,amssymb,
 aps,
 longbibliography,
 prb,
 floatfix
]{revtex4-2}

\usepackage{graphicx}

\usepackage[bb=dsserif]{mathalpha}

\usepackage[%
  colorlinks=true,
  urlcolor=blue,
  linkcolor=blue,
  citecolor=blue
]{hyperref}

\DeclareMathOperator{\tr}{tr}

\def\tcm{T.C.M. Group, Cavendish Laboratory, University of Cambridge, J.J. Thomson Avenue, Cambridge, CB3 0HE, UK}
\def\tauni{Raymond and Beverly Sackler School of Physics and Astronomy, Tel Aviv University, Tel Aviv 6997801, Israel}

\begin{document}

\title{Conductivity scaling and absence of localization in disordered nodal line semimetals}% Force line breaks with \\

\author{Carolina Paiva}
\affiliation{\tauni}

\author{Jan Behrends}
\affiliation{\tcm}

\begin{abstract}
Transport plays a key role in characterizing topological insulators and semimetals.
Understanding the effect of disorder is crucial to assess the robustness of experimental signatures for topology.
In this work, we find the absence of localization in nodal line semimetals for long-range scalar disorder and a large range of disorder strengths.
Using a continuum transfer matrix approach, we find that the conductivity in the plane and out of the plane of the nodal line increases with system size and disorder strength.
We substantiate these findings by a perturbative calculation and show that the conductivity increases with disorder strength using the Kubo formula in the self-consistent Born approximation.
We also find that the system remains metallic for vector disorder and that vector disorder can drive a transition from an insulating to a metallic regime.
Our results demonstrate the absence of localization in a three-dimensional bulk system.
\end{abstract}

\maketitle

\section{\label{sec:introduction}Introduction}

The scaling theory of localization~\cite{Abrahams:1979iv} postulates that the conductance of a disordered electron system at zero temperature scales universally with system size.
In a large class of systems of linear dimension $L$, the logarithmic derivative $\beta (g)= \mathrm{d} \log (g) /\mathrm{d} \log(L)$, with the dimensionless conductance $g=G/(e^2/h)$ and conductance $G$, depends only on the dimension of the system $d$ and on $g$ itself.
Quantum corrections play an important role in two dimensions (2D)~\cite{Evers:2008gi}:
Indeed while $\beta \approx 0$ for large $g$, the sign of $\beta (g \gg 1)$ is determined by the system's symmetry class~\cite{Hikami:1981eq}, resulting in a metal-insulator transition, localization at any disorder strength, or critical behavior~\cite{Gade:1991ka,Evers:2008gi}.

However, topological systems might defy this expectation.
A $\theta$-term~\cite{Pruisken:1984cj} drives 2D Dirac fermions into a metallic phase~\cite{Ostrovsky:2007hc} such that they evade localization for all disorder strengths~\cite{Bardarson:2007iu,Nomura:2007jb}.
Similarly, 3D Weyl nodes exhibit a phase transition or crossover~\cite{Nandkishore:2014cv,Pixley:2021do} from a pseudo-ballistic to a diffusive regime~\cite{Goswami:2011hh} with a positive $\beta$ function for all disorder strengths~\cite{Sbierski:2014bo,Syzranov:2017hw}, providing an example of a non-Anderson disorder-driven phase transition~\cite{Syzranov:2017hw,Zhu:2023hr}.

In this work, we numerically demonstrate that 3D nodal line semimetals subject to long-range disorder remain metallic, even for strong disorder.
Nodal line semimetals are topological semimetals where the bands touch in lines in the Brillouin zone~\cite{Heikkila:2011hw,Beri:2010dk,Burkov:2011ek}.
Different from Weyl nodes, additional symmetries are necessary to protect this gap closing~\cite{Beri:2010dk,Burkov:2011ek,Chiu:2014fi,Yang:2020he}.
Following theoretical proposals~\cite{Phillips:2014kj,Fang:2015gt,Chan:2016ho,Bzdusek:2017dy,Wang:2017dh,Behrends:2017cv}, nodal lines close to the Fermi energy have been realized in solid state systems~\cite{Bian:2016bn,Schoop:2016fv,Takane:2018fe,Chang:2019kz,Fu:2019gh,Lv:2021er}, photonic crystals~\cite{Deng:2019kj}, and cold atom systems~\cite{Song:2019gl}.

\begin{figure}
\includegraphics[scale=1]{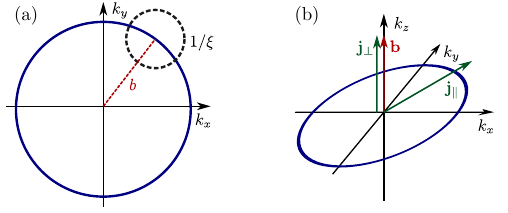}
\caption{(a) Nodal line with radius $b$ in the $x$-$y$ plane (orthogonal to $\mathbf{b}=b \mathbf{e}_z$ forming the Fermi surface (blue). The Gaussian scattering potential has range $1/\xi$ (dashed line). (b) We denote the conductivity in the plane orthogonal to the plane of the nodal line by $\sigma_\perp$ (current $\mathbf{j}_\perp$ in green) and in the plane of the nodal line by $\sigma_\parallel$ (current $\mathbf{j}_\parallel$ in green).}
\label{fig:nodal_line}
\end{figure}

Long-range disorder in nodal line semimetals may increase the conductivity due to weak anti-localization~\cite{Syzranov:2017dh,Chen:2019cz,Yang:2022ct}.
This is in stark contrast to short-range disorder that results in weak localization~\cite{Chen:2019cz}, and, via a transition from a semimetallic to a metallic regime, eventually to Anderson localization~\cite{Goncalves:2020ci}.
Here, short- and long-range disorder refers to the disorder correlation length $\xi$ compared with the only other length scale $1/b$, the inverse of the nodal line radius: Disorder is short-range when $b\xi \ll 1$ and long-range when $b\xi \gg 1$; cf.\ Fig.~\ref{fig:nodal_line}(a).
Weak anti-localization arises when the relevant scattering processes are effectively 2D, whereas the weak localization correction is a 3D effect~\cite{Chen:2019cz}.
Similarly to long-range disorder, a constant line broadening of the Green's function has been found to increase the conductivity~\cite{Mukherjee:2017fe}.
Recent magnetotransport experiments are consistent with weak anti-localization~\cite{An:2019ko,Laha:2019ei,Zhou:2020ex}.

While the weak anti-localization correction is an inherently perturbative result, the behavior beyond the perturbative regime remains elusive.
Here we use a nonperturbative transfer matrix approach~\cite{Bardarson:2007iu,Cheianov:2006hl,Titov:2007ev,Mello:1988cj,Tamura:1991ki,Beenakker:1997gz} to demonstrate the absence of localization for a nodal ring at zero energy in the presence of long-range scalar disorder, i.e., a smooth disorder potential with Gaussian correlations.
One advantage of the transfer matrix approach is the direct accessibility of the nodal point, which evades subtleties arising from different orders of limits in perturbative calculations.
We find that the conductivity increases with disorder and see no indication of localization for strong disorder.
While the initial growth of the conductivity with disorder could be expected since the density of states at zero energy grows with disorder, we find here a persistent absence of localization for a large range of disorder strengths.
For nodal lines whose diameter is larger than the inverse scattering length, we find that the conductivity collapses onto the same line, consistent with the 2D-dominated weak anti-localization regime~\cite{Syzranov:2017dh,Chen:2019cz}.

We focus on transport out of the plane of the nodal line [cf.\ Fig.~\ref{fig:nodal_line}(b)] and complement our results with a perturbative approach in the self-consistent Born approximation (SCBA), which also predicts an unbounded growth of the conductivity with disorder.
Furthermore, we show that these conclusions remain true for symmetry-breaking disorder, which can drive the system to a metallic phase, even when starting from a gapped insulator phase.
For in-plane transport, we find that the conductivity oscillates with system size in the clean case and that these oscillations remain visible upon the inclusion of disorder; for clean systems, the conductivity in the limit of large systems does not equal the Kubo formula result, different from out-of-plane transport.

The Fermi surface we consider, i.e., a 1D nodal line embedded in a 3D system, has codimension 2~\cite{Chiu:2014fi}, which, together with the linear dispersion, has led to interpretations of nodal line semimetals as a 3D analog of graphene~\cite{Bian:2016bn}.
The growth of the conductivity with disorder is analogous to the conductivity growth in graphene without intervalley scattering, albeit with one important difference:
The unbounded growth of the conductivity can be found only for isolated Dirac nodes~\cite{Bardarson:2007iu} as intervalley scattering will eventually lead to localization---the true absence of localization is thus only possible when isolated Dirac nodes arise as surface states of topological insulators~\cite{Nomura:2007jb}.
However, isolated nodal lines, as considered in this work, exist in the bulk of lattice systems, and hence the absence of localization does not rely on the absence of certain scattering processes, apart from coupling between bulk and surface states.
Although we obtain all of our results at zero chemical potential, we expect that, analog to the pseudodiffusive regime in graphene~\cite{Prada:2007im}, the conductivity we obtain here can be observed in measurements, either for small doping or as a minimal conductivity~\cite{Novoselov:2005es}.

Transport experiments are a crucial testbed for probing topological semimetals~\cite{Xiong:2015kl,Gooth:2017bn}, including the absence of localization we predict in this work.
Thanks to the growing number of material candidates that have been theoretically predicted or experimentally verified to host nodal lines~\cite{Xu2011_ex,Yu2015_ex,Kim2015_ex,Zhou:2020ex,Chen2015_ex}, we are confident that our predictions can be tested using suitable experimental platforms with a near-dispersionless nodal line close to the Fermi level.

This work is organized as follows:
We introduce the model for the nodal line semimetal and summarize the transfer matrix method in Sec.~\ref{sec:model}, where we also analytically derive the conductivity for the clean case.
In Sec.~\ref{sec:disorder}, we introduce correlated scalar disorder and present the numerically computed conductivity out of the plane of the nodal line.
After comparing these findings with a perturbative approach in Sec.~\ref{sec:Vertex_Correction}, we demonstrate that the metallic regime remains stable for symmetry-breaking disorder in Sec.~\ref{sec:vector_potential}.
We briefly discuss in-plane transport in Sec.~\ref{sec:inplane} before concluding and discussing potential implications in Sec.~\ref{sec:conclusion}.

\section{\label{sec:model}Model and transport in clean systems}

%We utilize transfer and scattering matrices to compute the conductance of a disordered nodal line semimetal.
We start by considering finite systems and employ the scattering theory of electronic conduction~\cite{Buttinger:1985ib}, using transfer and scattering matrices to compute the conductance of a disordered nodal line semimetal within the Landauer framework.
To this end, we consider a circular nodal line at the chemical potential $\mu=0$, described by the continuum Hamiltonian~\cite{Burkov:2011ek}
\begin{equation}
  \mathcal{H} = \hbar v \sum_{j=1}^{3} (\hat{k}_j\Gamma_{j} + b_{j}\Gamma_{j,5}) + m\Gamma_{4} + V (\mathbf{r})
\label{eq:HamiltonianNodalLine}
\end{equation}
with velocity $v$, momentum operator $\hat{\mathbf{k}}$, and the indices $1,2,3$ denote the spatial directions $x,y,z$.
Here we consider a scalar disorder potential $V(\mathbf{r}) = u (\mathbf{r}) \mathbb{1}$ with the identity matrix $\mathbb{1}$
(we discuss a different type of disorder with a vector potential in Sec.~\ref{sec:vector_potential}).
The matrices $\Gamma_{1 \dots 4}$ are the Euclidean gamma matrices, i.e., $4\times 4$ Hermitian anticommuting matrices satisfying $\{\Gamma_{i},\Gamma_{j}\} = 2\delta_{ij}$. Their product $\Gamma_5 = \Gamma_1 \Gamma_2 \Gamma_3 \Gamma_4$ is also Hermitian, and we define $\Gamma_{i,j} = (i/2) [\Gamma_{i},\Gamma_{j}]$ with $i<j$ for convenience.
In the clean case ($u(\mathbf{r})=0)$, the Hamiltonian describes the low-energy degrees of freedom of a nodal line semimetal when $\hbar v |\mathbf{b}|>m$:
the spectrum is then gapless with a circular nodal line with radius $\sqrt{|\mathbf{b}|^2-(m/(\hbar v))^2}$ in the plane orthogonal to the vector $\mathbf{b}$; when $\hbar v |\mathbf{b}| < m$, a gap opens in the spectrum~\cite{Burkov:2011ek}.

To calculate the conductance at the nodal line along the $z$ direction, we switch to real space along $z$ and seek for zero-energy solution of the Hamiltonian ($\mathcal{H}\psi= 0$), given by
\begin{equation}
\frac{d\psi}{dz} = -i\Gamma_{3} \left( \sum_{j=1}^2 k_j \Gamma_j + \sum_{j=1}^3 b_{j}\Gamma_{j,5} + \frac{m}{\hbar v} \Gamma_{4} + \frac{u(\mathbf{r})}{\hbar v} \right) \psi .
\end{equation}
This equation can be formally solved by
\begin{equation}
    \psi_{\mathbf{k}_{\perp}}(z) = T_{\mathbf{k}_{\perp},\mathbf{k}'_{\perp}}(z,z')\psi_{\mathbf{k}'_{\perp}}(z')
\end{equation}
with the transfer matrix
\begin{widetext}
\begin{align}
T_{\mathbf{k}_{\perp},\mathbf{k}_{\perp}'}(z,z') =& \mathcal{P} \exp \left\{ -i\Gamma_3 \int_{z'}^{z} dz'' \left[ \left( k_x \Gamma_1 + k_y \Gamma_2 + \frac{m}{\hbar v} \Gamma_{4} + \sum_{j} b_{j}\Gamma_{j,5} \right) \delta_{\mathbf{k}_\perp,\mathbf{k}_\perp'} + \frac{u_{\mathbf{k}_\perp-\mathbf{k}_\perp'} (z'') \mathbb{1}}{\hbar v} \right] \right\}
\label{eq:transfer_matrix}
\end{align}
\end{widetext}
where $\mathcal{P}$ denotes path ordering and $u_{\mathbf{q}_\perp}(z)$ is the Fourier transform of $u(\mathbf{r})$ along the $x$ and $y$ directions.
The transfer matrix $T$ is alternatively described by a corresponding scattering matrix $S = \begin{pmatrix} r & t' \\ t & r' \end{pmatrix}$ with transmission matrices $t,t'$ and reflection matrices $r,r'$~\cite{Mello:1988cj,Beenakker:1997gz}.
The dimensionless conductance $g = \tr[ t^\dagger t]$ can directly be computed from the transmission matrix via the Landauer formula~\cite{Buttinger:1985ib}.

We pause here to note that we generally need to resort to numerical evaluations of Eq.~\eqref{eq:transfer_matrix} for disordered systems, as described in detail in Sec.~\ref{sec:disorder}.
In the clean case, however, since the transfer matrix is diagonal in momentum space, the path-ordered exponential equals the matrix exponential.
We can find analytical solutions for transport parallel to $\mathbf{b}$, i.e., $\mathbf{b} = b \mathbf{e}_z$.
For simplicity, we focus on $m=0$ and discuss the solution for $m\neq 0$ in Appendix~\ref{sec:mass_term}.
When $m=0$, the system is effectively time-reversal symmetric; we find~\cite{gamma_conv} $\mathcal{H}= \Gamma_2 \mathcal{H}^* \Gamma_2 $ for complex $\Gamma_2$ and real $\Gamma_{1}$ and $\Gamma_{3}$.

\begin{figure}
\includegraphics{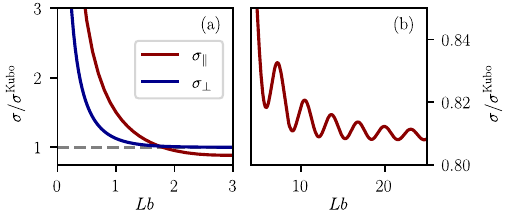}
\caption{Conductivity in the clean case for transport out of the plane ($\sigma_\perp$, blue) and in the plane of the nodal line ($\sigma_\parallel$, dark red), divided by the Kubo formula result $\sigma_\perp^{\mathrm{Kubo}} = b/\pi$ and $\sigma_\parallel^{\mathrm{Kubo}} = b/(2\pi)$. Panel (a) shows the exponential decay for short $Lb$, and panel (b) the oscillations of $\sigma_\parallel$ for large $L b$.}
    \label{fig:Numerical_Integral_PlaneNodalLine}
\end{figure}

When choosing $\Gamma_3 = \tau_3 \otimes \sigma_0$, the transfer matrix has the form $T (k_x,k_y) = \begin{pmatrix} [t^\dagger]^{-1} & r' {t'}^{-1} \\ -{t'}^{-1} r & {t'}^{-1} \end{pmatrix}$, thus, its diagonal $2 \times 2$ blocks equal $[t^\dagger]^{-1}$ and ${t'}^{-1}$~\cite{Mello:1988cj}.
The transmission matrix for a system of length $L_z = L$ along the transport direction and width $L_x = L_y = W$ is then
\begin{equation}
    t = \begin{pmatrix}
    \frac{2 \cosh(bL) \cosh(kL)}{\cosh (2 b L) + \cosh (2 k L)} &
    \frac{2 e^{-i \phi} \sinh(bL) \sinh(kL)}{\cosh (2 b L) + \cosh (2 k L)} \\
    \frac{2 e^{ i \phi} \sinh(bL) \sinh(kL)}{\cosh (2 b L) + \cosh (2 k L)} & 
    \frac{2 \cosh(bL) \cosh(kL)}{\cosh (2 b L) + \cosh (2 k L)}
    \end{pmatrix},
\end{equation}
with momenta $k_x  = 2\pi n_x/W$, $k_y = 2\pi n_y/W$ and polar coordinates $k_x = k \cos (\phi)$, $k_y = k\sin (\phi)$.
In the limit $W \gg L$, the dimensionless conductance
\begin{equation}
 g_\perp = \frac{W^2}{2\pi} \int d k k \tr [ t^\dagger t] = \frac{W^2}{2\pi}\frac{\log[ 4 \cosh^2(b L) ]}{L^2} .
\label{eq:ConductanceOutofPlane}
\end{equation}
We will henceforth focus on the conductivity $\sigma_\perp = g_\perp L/W^2$, which we plot for the clean case in Fig.~\ref{fig:Numerical_Integral_PlaneNodalLine}.
For large systems with $L \gg 1/b$, we recover the Kubo formula result $\sigma_\perp^\mathrm{Kubo} = b/\pi$~\cite{Burkov:2011ek}, which we will revisit in Sec.~\ref{sec:Vertex_Correction}.
For mesoscopic systems, however, the Kubo formula is not necessarily recovered~\cite{Nikolic:2001cf}.

Thus far we have explored the out-of-plane transport in the clean limit.
To investigate transport in the plane of the nodal line, we change $\mathbf{b} \to b \mathbf{e}_x$ and compute the transmission matrix $t$ analog as above, again for $W\gg L$, giving the dimensionless conductance~\cite{JanPhD}
\begin{equation}
g_\parallel = \frac{W^2}{\pi^2} \int \frac{d^2 k}{1 + \cos q_+ \cos q_- + \frac{L^2 (b^2 + k^2)}{ q_+ q_-} \sin q_+ \sin q_-}
\label{eq:Conductance_Parallel_Transport}
\end{equation}
with $q_\pm = L \sqrt{b^2 - k^2 \pm 2i b k \cos \phi}$.
Since we cannot find an analytical solution of this integral, we integrate~\eqref{eq:Conductance_Parallel_Transport} numerically and show the resulting conductivity $\sigma_\parallel = g_\parallel L/W^2$ in Fig.~\ref{fig:Numerical_Integral_PlaneNodalLine}.
Two features are noteworthy: The conductivity oscillates with $L$ with period $\pi/b$, Fig.~\ref{fig:Numerical_Integral_PlaneNodalLine}(b).
The oscillations vanish for large $Lb \gg 1$ (the decay is well-approximated by $\sim 1/(L b)^{3/2}$~\footnote{We thank Jun-Won Rhim for this observation.}), such that $\sigma_\parallel$ approaches a constant for large systems $L \gg 1/b$.
Remarkably, this constant does not equal the Kubo formula result, $\sigma_\parallel^{\mathrm{Kubo}} = b /(2\pi)$, but approximately $\sigma_\parallel/\sigma_\parallel^{\mathrm{Kubo}} \approx 0.81$.

\section{\label{sec:disorder}Out-of-plane transport in disordered systems}

We now turn to disordered systems.
To this end, we consider long-range disorder with $\langle u(\mathbf{r}) \rangle_\mathrm{dis} = 0$ and correlations
\begin{equation}
    \langle u(\mathbf{r}) u(\mathbf{r'})\rangle_\mathrm{dis} = \frac{K_0}{(2\pi)^{3/2}}\left(\frac{\hbar v}{\xi}\right)^2 \exp\left(-\frac{(\mathbf{r}-\mathbf{r}')^2}{2\xi ^2}\right)
    \label{eq:CorrelationPotential}
\end{equation}
where $\langle \dots \rangle_\mathrm{dis}$ denotes the disorder average. The disorder potential is characterized by the dimensionless disorder strength $K_0$ and the correlation length $\xi$. Individual disorder realizations are typically smooth and vary on the scale of $\xi$.

\begin{figure}
\includegraphics{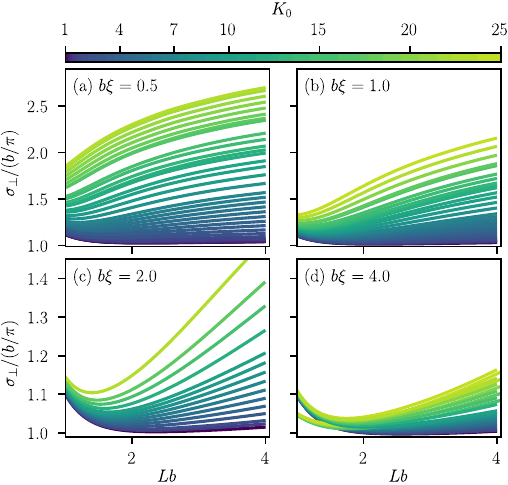}
\caption{Conductivity of a disordered nodal line semimetal at $\mu=0$ as a function of length along the transport direction $L$.
The different panels show different values of $b\xi$ and different colors denote different disorder strengths $K_0$.
For sufficiently large $Lb \gtrsim 2$, the conductivity increases with $L$ and $K_0$ for all values of $b\xi$ for all disorder strengths we consider.
All results are for $W=20/b$ and $M_x=M_y=19$ and averaged over up to 300 disorder realizations.}
\label{fig:conductivity_raw}
\end{figure}

Since the disorder potential is smooth, we can ignore the path-ordering in the transfer matrix $T_{\mathbf{k}_\perp,\mathbf{k}_\perp'} (z,z')$ [Eq.~\eqref{eq:transfer_matrix}] when $z-z' \ll \xi$. We can thus split up the transfer matrix
\begin{equation}
 T (L,0) \approx T (z_N,z_{N-1}) \dots T (z_1,z_0)
\end{equation}
into a finite product of transfer matrices for slices with $z_0 = 0$, $z_N = L$, and $z_{j+1}-z_j \ll \xi$, which we evaluate numerically~\cite{Bardarson:2007iu,Cheianov:2006hl,Titov:2007ev}. The equality is exact when $N\to\infty$.
For the numerical evaluation, we work in the limit $W \gg L$ in which $\sigma_\perp$ depends only on $L$, but not on $W$.

To work with finite-size matrices, we introduce a momentum cutoff $|k_j| \le k_j^\mathrm{max} =2\pi M/W$ with integer $M$ and $j=x,y$.
Unlike in graphene~\cite{Bardarson:2007iu} or Weyl semimetals~\cite{Sbierski:2014bo}, we cannot compute the transmission for arbitrary large system widths $W$ since the momentum cutoff needs to be larger than the nodal line radius, $2\pi M/W > b$; since $M$ is limited by computational resources, $W b$ (and hence $Lb$ for fixed $W/L$) is limited to a maximal value as well.

We show the ensemble-averaged conductivity $\sigma_\perp$ divided by $b/\pi$ as a function of $L$ in Fig.~\ref{fig:conductivity_raw}.
The color denotes the disorder strength $K_0$ and different panels show different values of $b\xi$.

We find that the conductivity increases with the strength of the disorder potential.
This increase is stronger for small values of $b\xi$.
These results are compatible with the behavior described in Refs.~\onlinecite{Syzranov:2017dh,Chen:2019cz}, where it was argued that long-range impurities lead to weak anti-localization, i.e., an enhanced conductivity compared with the clean case~\cite{Syzranov:2017dh,Chen:2019cz}. 
Here, going beyond the perturbative regime, we find that the system remains metallic for a wide range of disorder strengths with no indication of a transition to an insulator for large $K_0$. 
We stress that we expect this behavior to hold only for long-range disorder:
For short-range (uncorrelated white noise) disorder, previous works found weak localization, i.e., a decreasing conductivity~\cite{Chen:2019cz}, and eventually a transition to an Anderson-localized regime for strong disorder~\cite{Goncalves:2020ci}.
Although, as we would like to emphasize here, we cannot infer a transition to localization from our results for small $b\xi$, this is not in disagreement with these previous studies since we can neither access arbitrarily small $b\xi$ nor large systems.
At very large $Lb$, a similar transition to weak localization with a fixed conductivity could occur, as we explore perturbatively in Sec.~\ref{sec:Vertex_Correction}.

The crossover from weak anti-localization for long-range disorder to weak localization for short-range disorder can be understood as a crossover from 2D diffusion to 3D diffusion~\cite{Chen:2019cz}:
For $\mu\neq 0$, the Fermi surface is effectively a torus.
When the momentum-space range of the scattering potential $1/\xi < b$, backscattering is effectively limited to opposite poloidal directions, i.e., backscattering occurs within 2D planes along a fixed toroidal direction of the torus.
In contrast, when $1/\xi>b$, backscattering can occur between opposite toroidal directions, leading to a 3D diffusive behavior~\cite{Chen:2019cz}.
Our results indicate that the 2D diffusion also holds for $\mu=0$.

\begin{figure}
\includegraphics{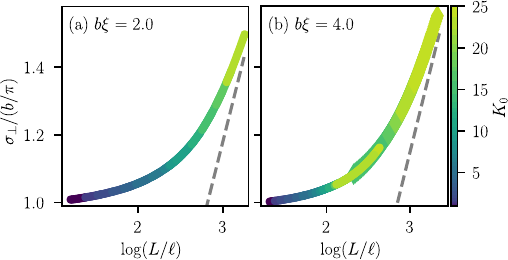}
\caption{Rescaled conductivity of a disordered nodal line semimetal, data from Fig.~\ref{fig:conductivity_raw} for $L \gtrsim l_\mathrm{mfp}$.
For sufficiently large $b\xi$ [$b\xi=2$ in panel (a) and $b\xi=4$ in panel (b)], we can find a length $\ell$ such that $\sigma_\perp$ follows the same scaling for all $K_0$ as a smooth function of $L/\ell$.
Triangles in (b) are for $W=40/b$ ($Lb \le 8$), discs for $W=20/b$ ($Lb \le 4$).
Both curves collapse for sufficiently large $Lb$.
The dashed lines denote the asymptotic $\sigma_\perp = b/\pi \log(L/\ell)$ scaling.}
\label{fig:conductivity_rescaled}
\end{figure}

We verify the 2D behavior by a one-parameter scaling of the conductivity:
The number of states at the Fermi level is $n \sim b W$, which independently contribute to the dimensionless conductance $g = n g_\mathrm{2D}$ where $g_\mathrm{2D}$ is the 2D conductance $g_\mathrm{2D} = \sigma_\mathrm{2D} W/L$.
2D scaling implies that $\sigma_\mathrm{2D}$ follows a one-parameter scaling for large $L >1/b$ and $L \gtrsim l_\mathrm{mfp}$, where $l_\mathrm{mfp}$ is the mean free path.
Since the system is at $m=0$ in symmetry class AII, we expect $\sigma_\mathrm{2D} = (1/\pi) \log(L/\ell)$~\cite{Evers:2008gi} to hold for large values of $\sigma_\mathrm{2D}$.
Then the 3D conductivity $\sigma_\perp = b/\pi \log (L/\ell)$ where the length $\ell$ depends on the disorder strength and on $b\xi$.
For sufficiently large $b\xi$, the conductivity $\sigma_\perp/(b/\pi)$ collapses onto the same line, which we show in Fig.~\ref{fig:conductivity_rescaled}.
We find a smooth scaling for $b\xi=2$ [panel (a)] and $b\xi=4$ [panel (b)].

\subsection{\label{sec:Vertex_Correction}Diagramatic approach: Vertex correction}

In this section, we compute the conductivity diagrammatically using the self-consistent Born approximation (SCBA) to complement the exact transfer matrix results from the Landauer formalism.
While the transfer-matrix approach is well-suited to study the 2D-diffusive regime and the crossover to 3D diffusion, the diagrammatic calculation gives the 3D results in the limit of large system sizes, $L\gg 1/b$ and $L\gg \xi$.
We follow previous approaches carried out for graphene~\cite{Noro:2010ct} and Weyl semimetals~\cite{Ominato:2014ch}, but need to resort to numerically evaluated expressions at early stages of the calculation due to the anisotropy of the model.
We briefly sketch the derivation for the nodal line semimetal within the SCBA before presenting the results.
Using the SCBA and vertex correction, we neglect all crossed diagrams, which give the weak localization correction~\cite{RammerBook}. Although the diagrammatic calculation is thus only complementary to the exact transfer matrix method, we find in this section that the general trends are reproduced due to suppression of crossed diagrams for long-range disorder.

The starting point is the Hamiltonian~\eqref{eq:HamiltonianNodalLine}, which, for $m= V(\mathbf{r})=0$, can, by a momentum-dependent unitary transformation $\mathcal{H}_\mathbf{k}' = \mathcal{U}_\mathbf{k} \mathcal{H}_\mathbf{k} \mathcal{U}_\mathbf{k}^\dagger$, be brought into a block-diagonal form (using the convention~\cite{gamma_conv})
\begin{align}
 \mathcal{H}_\mathbf{k}' = \begin{pmatrix}
       \mathcal{H}_{-} & 0\\
       0 & \mathcal{H}_{+}
    \end{pmatrix},
  & & \mathcal{U}_\mathbf{k} = e^{\frac{i\pi}{4} (\tau_1\sigma_3- \tau_2\sigma_2 )} e^{\frac{i}{2} (\pi \tau_3 + \phi \sigma_3)}
\label{eq:BlockDiagonalHamiltonian}
\end{align}
with $\mathcal{H}_{\pm} = (k \pm b)\sigma_{y}+k_{z}\sigma_{x}$ and $k = \sqrt{k_x^2+k_y^2}$ in polar coordinates.
When $b>0$, $\mathcal{H}_+$ is gapped and $\mathcal{H}_{-}$ is gapless; note that $\mathcal{H}_-$ has been previously considered as an effective Hamiltonian for a nodal line semimetal~\cite{Rhim:2016dy}.
Disorder couples both blocks since the transformed potential $V_{\mathbf{k}-\mathbf{k'}}' = \mathcal{U}_\mathbf{k} V_{\mathbf{k}-\mathbf{k'}}\mathcal{U}_\mathbf{k'}^\dagger$ is generally off-diagonal, e.g., when $V_{\mathbf{k}-\mathbf{k'}} = u_{\mathbf{k}-\mathbf{k}'} \mathbb{1}$, we have $V_{\mathbf{k}-\mathbf{k'}} \to V_{\mathbf{k}-\mathbf{k'}}' = u_{\mathbf{k}-\mathbf{k'}} \mathcal{Q} (\phi-\phi')$ with
\begin{align}
 \mathcal{Q} (\phi) = \begin{pmatrix} \cos (\phi/2) \sigma_0  & -i\sin (\phi/2) \sigma_{x} \\ -i\sin (\phi/2) \sigma_{x} &  \cos (\phi/2)  \sigma_0 \end{pmatrix}
\end{align}

\begin{figure}
    \includegraphics{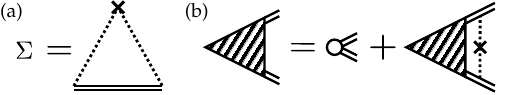}
    \caption{(a) Diagrammatic representation of the self-energy correction in the SCBA. The double solid line represents the disorder-averaged Green's function and the dashed lines disorder scattering. (b) Diagrammatic representation of the vertex correction. The triangle represents the velocity-vertex operator and the circle the bare velocity operator (here: $\sigma_x$).}
    \label{fig:diagrams}
\end{figure}

With the Hamiltonian~\eqref{eq:BlockDiagonalHamiltonian}, we can iteratively compute the disorder-averaged Green's function at zero energy in the SCBA, which for the $n$th iteration reads
\begin{equation}
 G^{s}_{n}(\mathbf{k}) = \frac{1}{-\mathcal{H}_\mathbf{k}'-\Sigma^{s}_{n}(\mathbf{k})}
 \label{eq:GreensFunction}
\end{equation}
with $s=1$ for the retarded and $s=-1$ for the advanced Green's function.
In the SCBA the self-energy 
\begin{equation}
    \Sigma^{s}_{n}(\mathbf{k}) = \sum_{\mathbf{k'}} \langle | u_{\mathbf{k}-\mathbf{k'}}|^2 \rangle_\mathrm{dis} \mathcal{Q}(\phi-\phi') G^{s}_{n-1}(\mathbf{k'}) \mathcal{Q} 	(\phi'-\phi) ,
    \label{eq:SelfEnergy}
\end{equation}
as diagrammatically represented in Fig.~\ref{fig:diagrams}(a).
Eqs.~\eqref{eq:GreensFunction} and~\eqref{eq:SelfEnergy} give coupled, self-consistent equations, with the disorder average $\langle | u_{\mathbf{q}} |^2 \rangle_\mathrm{dis}$ from~\eqref{eq:CorrelationPotential} and the self-energy $\Sigma^{s}_{0} = -is\eta \sigma_{0}\tau_{0}$ in the initial iteration, where $\eta>0$ is a small parameter chosen such that it does not affect the results for $n \gg 1$.

Since $G^{s}_{n}(\mathbf{k})$ depends only on the cylindrical coordinates $k$ and $k_{z}$ but not on the angle $\phi$, we can perform the integration over $\phi'$ in Eq.~\eqref{eq:SelfEnergy} without prior knowledge of the Green's function itself.
For scalar disorder, the Green's function and self-energy remain block-diagonal for all SCBA iterations, namely
\begin{align}
 \Sigma^{s}_{n}(\mathbf{k}) =& \frac{K_{0}\xi}{2} \left( \frac{\hbar v}{2\pi} \right)^2 \int dk_z' dk k' e^{-\frac{\xi^2}{2} \left[ (k_{z}-k_{z}')^2 + (k^2 + {k'}^2) \right]} \nonumber  \\
& \times \left( I_0 ( \xi^2 k k') \tilde{G}^{s,+}_{n-1}(\mathbf{k'}) + I_1 ( \xi^2 k k') \tilde{G}^{s,-}_{n-1}(\mathbf{k'}) \right)
\end{align}
with the modified Bessel functions of the first kind $I_0(x)$ and $I_1(x)$, and the block-diagonal
\begin{align}
  \tilde{G}^{s,\pm}_{n}(\mathbf{k})
  &= G^s_{n} (\mathbf{k}) \pm \tau_x \sigma_x G^s_{n} (\mathbf{k}) \sigma_x\tau_x \\
  &= \begin{pmatrix} g_{n,\mathbf{k}}^{s,-} \pm \sigma_x g_{n,\mathbf{k}}^{s,+} \sigma_x & \\ & g_{n,\mathbf{k}}^{s,+} \pm \sigma_x g_{n,\mathbf{k}}^{s,-} \sigma_x \end{pmatrix},
\end{align}
where the $g_{n,\mathbf{k}}^{s,\mp}$ are the upper and lower blocks of the Green's function
\begin{equation}
 G_{n}^s (\mathbf{k}) = \begin{pmatrix} g_{n,\mathbf{k}}^{s,-} & \\ & g_{n,\mathbf{k}}^{s,+} \end{pmatrix}.
\end{equation}

We now turn to the velocity-vertex operator, diagrammatically represented in Fig.~\ref{fig:diagrams}(b).
For transport along $z$, the bare velocity operator $J_0^{ss'} = (1/\hbar v) \partial_{k_z} \mathcal{H}_{\mathbf{k}}' =\sigma_{x}$, giving the velocity-vertex operator~\cite{Noro:2010ct,Ominato:2014ch}
\begin{equation}
    J^{ss'}_{n}(\mathbf{k}) = \sigma_{x} + \sum_{\mathbf{k'}} \langle V_{\mathbf{k}-\mathbf{k'}}' G^{s}(\mathbf{k'})J^{ss'}_{n-1}(\mathbf{k'})G^{s'}(\mathbf{k'}) V_{\mathbf{k'}-\mathbf{k}}' \rangle_\mathrm{dis}
\end{equation}
where $G^s (\mathbf{k}) = \lim_{n\to \infty} G_n^s (\mathbf{k})$ is the disorder-averaged Green's function in the SCBA approximation.
Here we again perform the integration over $\phi'$ analytically before numerically integrating over $k'$ and $k_{z}'$.

\begin{figure}
\includegraphics{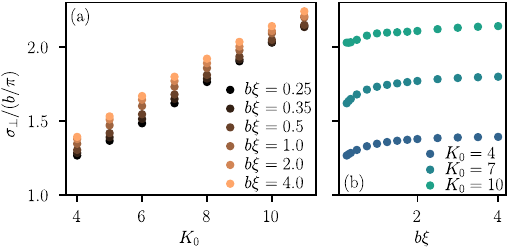}
\caption{Out-of-plane conductivity $\sigma_\perp$ computed via the vertex correction, Eq.~\eqref{eq:vertex_correction}.
(a) Conductivity as a function of disorder strength $K_0$ with colors denoting $b\xi$.
(b) Conductivity as a function of $b\xi$ with colors denoting $K_0$.}
\label{fig:vertex}
\end{figure}

Finally, the conductivity divided by $e^2/h$~\cite{Noro:2010ct,Ominato:2014ch}
\begin{equation}
 \sigma_{\perp} = -\frac{(\hbar v)^2}{2}\sum_{s,s'}ss' \int \frac{d^3\mathbf{k}}{(2\pi)^3} \tr[\sigma_{x}G^{s}(\mathbf{k})J^{ss'}(\mathbf{k})G^{s'}(\mathbf{k})]
 \label{eq:vertex_correction}
\end{equation}
where $J^{ss'} (\mathbf{k}) = \lim_{n\to \infty} J_n^{ss'} (\mathbf{k})$ is the disorder-averaged velocity-vertex operator in the SCBA approximation.
We show $\sigma_{\perp}$ in Fig.~\eqref{fig:vertex}.

We find that the conductivity increases with $K_0$ [Fig.~\eqref{fig:vertex}(a)], which supports the findings we obtained with the transfer matrix method presented in Sec.~\ref{sec:disorder}.
Furthermore, $\sigma_\perp/b$ increases with $b\xi$ and for $b\xi \gtrsim 1$, $\sigma_\perp/b$ approaches a $b\xi$-independent constant, which is consistent with $\sigma_\perp/b \propto \sqrt{(\Gamma/b)^2 + 1}$ found previously for a momentum- and energy-independent level-broadening $\Gamma$~\cite{Mukherjee:2017fe}.
Since we find that $\sigma_\perp/b$ is $b\xi$-independent for $b\xi \gtrsim 1$, our results indicate that $\Gamma/b$ depends only on $K_0$ in this regime.

\begin{figure}
\includegraphics[scale=1]{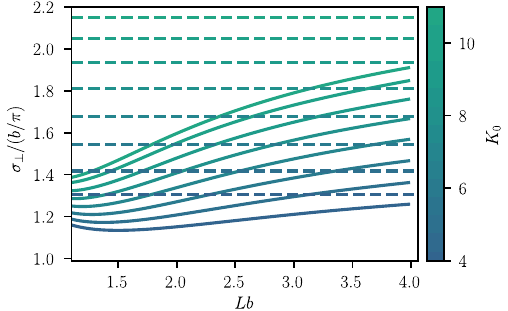}
\caption{Comparison of transfer matrix results (solid lines) and diagrammatic results (horizontal dashed lines) for $b\xi = 0.5$. The diagrammatic results are valid in the limit of large system size $L \gg \xi$ and $L \gg 1/b$, i.e., cannot reproduce the finite-size conductivity scaling captured by the transfer matrix approach.}
\label{fig:comparison}
\end{figure}

The conductivity we obtain via the transfer matrix method does not saturate for the system sizes we can achieve numerically.
In Fig.~\ref{fig:comparison}, we compare the diagrammatic SCBA results with (exact for sufficiently large $M$) transfer matrix results for finite $Lb$ and $b\xi=0.5$.
Although we do not achieve convergence with $Lb$, the transfer matrix results approach a constant, consistent with previous results~\cite{Chen:2019cz} that find a constant conductivity for a short-range scattering potential.
For larger values of $b\xi$, the diagrammatically obtained conductivity is much larger than the finite-size conductivity obtained using transfer matrices, which suggests that the $\sigma_\perp \propto \log (L/\ell)$ scaling holds for large $L/\ell$.

\subsection{\label{sec:vector_potential}Symmetry-breaking vector potential}

In this section, we demonstrate that the system remains metallic in the presence of disorder that breaks the symmetry protecting the nodal line.
Adding a term $u_0 \Gamma_{5}$ to the Hamiltonian~\eqref{eq:HamiltonianNodalLine} opens a gap:
$u_0 \Gamma_{5}$ breaks both the effective time-reversal symmetry, and inversion symmetry, which protects the nodal line.
Unlike other inversion-symmetry-breaking terms that give rise to a Weyl semimetal phase~\cite{Burkov:2011ek}, $u_0 \Gamma_{5}$ opens a gap everywhere in the spectrum and the system becomes insulating.

\begin{figure}
\includegraphics{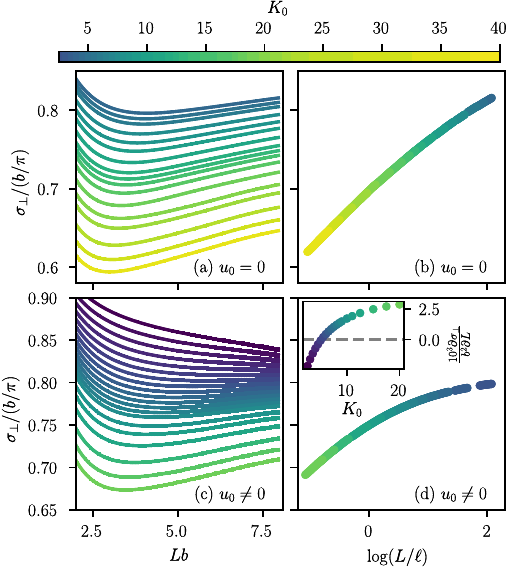}
\caption{Conductivity $\sigma_\perp$ for vector disorder of the form $V(\mathbf{r}) = u(\mathbf{r}) \Gamma_5$.
(a) Conductivity for $u_0=0$, $b\xi=2$ and various disorder strengths up to $K=40$ as a function of $L b$ and (b) as a function of $\log (L/\ell)$ for $L \gtrsim  l_\mathrm{mfp}$.
(c) Conductivity for $u_0=0.1 \hbar v/\xi$, $b\xi=2$ and various disorder strengths as a function of $L b$ and (d) for as a function of $\log (L/\ell)$ for $K_0 \ge 6$. The inset in (d) shows the slope $\partial \sigma_\perp/\partial L$ of linear fit for $7\le Lb \le 8$.
}
\label{fig:vector_disorder}
\end{figure}

To demonstrate the stability of the metallic phase towards symmetry-breaking disorder, we now consider a disorder potential of the form $V(\mathbf{r}) = u(\mathbf{r}) \Gamma_5$.
In Fig.~\ref{fig:vector_disorder}(a), we show numerical results for $u_0 \equiv \langle u(\mathbf{r}) \rangle_\mathrm{dis} = 0$ and correlated disorder with strength $K_0$ and range $\xi$ [Eq.~\eqref{eq:CorrelationPotential}] for a wide range of disorder strengths $K_0$ and fixed $\xi b = 2$.
We find that, although the conductivity itself decreases with $K_0$, the system remains metallic for all $K_0$ we consider.
We cannot exclude a transition to an insulating phase for larger values of $K_0>40$.

The symmetry-breaking disorder $\propto \Gamma_5$ may also drive a transition from an insulating to a metallic regime:
In Fig.~\ref{fig:vector_disorder}(c), we show numerical results for $u_0 \neq 0$ and $b \xi = 2$.
In the absence of disorder, i.e., for $\langle u (\mathbf{r}) u (\mathbf{r'})\rangle_c = \langle u (\mathbf{r}) u (\mathbf{r'})\rangle_\mathrm{dis} - u_0^2 = 0$, the system is gapped and therefore insulating.
For weak disorder, the conductivity is initially nonzero but decreases with system size.
For strong disorder $K_0 > K_c$ with $K_c = 5.25 \pm 0.5$, the conductivity increases with system size and follows a one-parameter scaling, cf.\ Fig.~\ref{fig:vector_disorder}(d).
Different from the case of scalar disorder, the conductivity does not grow indefinitely but flows towards a threshold value $\sigma_\perp/(b/\pi) = 0.8 \pm 0.05$.
(The data collapse for a rescaled $L$ is not possible for $K_0 < K_c$.)
As we show in the inset in Fig.~\ref{fig:vector_disorder}(d), the conductivity $\sigma_\perp$ decreases with $L$ below the threshold and increases with $L$ above the threshold, indicating a metal-insulator transition at $K_c$.
Since the conductivity for $K_0 < K_c$ is always $\sigma_\perp/(b/\pi) \gtrsim 0.8$, an alternative scenario is a flow towards the same threshold value for $K_0<K_c$.
A more detailed analysis is necessary to determine whether the system always tends towards an insulator below the threshold.

We note that a similar behavior has been observed for 2D class-D superconductors~\cite{Medvedyeva:2010jf} and weak topological insulator surface states~\cite{Mong:2012du} with a random mass term, which at the transition between two distinct insulators (corresponding to $\langle u (\mathbf{r}) \rangle_\mathrm{dis} =0$ here) remain metallic when disorder is added~\footnote{In class D, the system only becomes metallic for white-noise disorder~\cite{Medvedyeva:2010jf,Bardarson:2010jn}.}, and which away from the transition, i.e., starting from an insulating phase, are driven to a metallic phase.
Note that details of the phase diagrams in classes D~\cite{Medvedyeva:2010jf} and AII~\cite{Mong:2012du} differ around the transition line.
We leave the investigation of these details of the phase diagrams in 3D nodal line semimetals for future works.

\section{\label{sec:inplane}In-plane transport}

\begin{figure}
\includegraphics{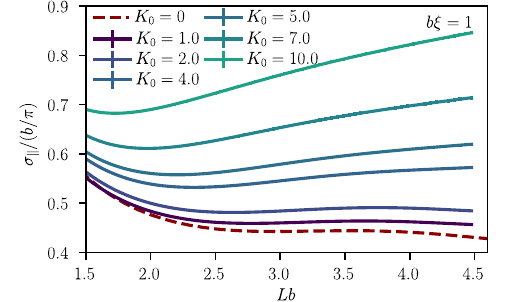}
\caption{Conductivity $\sigma_\parallel$ of a disordered nodal line semimetal as a function of the length along the transport direction $L$ with $\mathbf{b}$ orthogonal to the transport direction, i.e., transport in the plane of the nodal line.
Colors denote different disorder strengths $K_0$. Apart from oscillations pronounced for small $K_0$, the conductivity increases with $L$ and $K_0$ for all disorder strengths we consider.
Results are averaged over up to 200 disorder realizations.}
\label{fig:conductivity_ortho}
\end{figure}

We now turn to transport in the plane of the nodal line in the presence of disorder.
In Fig.~\ref{fig:conductivity_ortho}, we show the conductivity $\sigma_\parallel$ obtained using the transfer matrix approach as a function of $L$ for various values of $K_0$ and fixed $b\xi = 1$.
We also included the $K_0=0$ results [cf.\ Fig.~\ref{fig:Numerical_Integral_PlaneNodalLine}(b)] to show the clean-case oscillations as a function of $Lb$.

For weak disorder, the oscillations in $Lb$ remain stable until they become indiscernible from the increasing trend when $K_0 \sim 7$.
Analog to transport out of the plane of the nodal line, we see that the conductivity generally increases with $K_0$.
We expect that rescaling the conductivity similar to the data shown in Fig.~\ref{fig:conductivity_rescaled} is possible for sufficiently large system sizes when the oscillations do not play a role, although we did not attempt such a rescaling here.

\section{\label{sec:conclusion}Conclusion and Outlook}

In this work, we used a transfer matrix approach to show that the conductivity of a nodal line semimetal increases with disorder and that the system remains metallic for all disorder strengths investigated.
This holds for both transport directions, i.e., orthogonal to, and in the plane of the nodal line.
The metallic behavior is also sustained by the results obtained using the Kubo formula in the self-consistent Born approximation.
The system remains metallic even in the presence of vector disorder that breaks inversion symmetry, which crucially protects the nodal line; such correlated vector disorder can also drive a transition from an insulator to a metal.

We found oscillations in the conductivity for transport in the plane of the nodal line, which persist for weak disorder.
These oscillations decay with system size, such that, in the clean case, the conductivity goes to a constant for large systems.
However, the transfer matrix result does not match the Kubo formula result.

In real materials, various effects might play a role that we did not consider in this work:
First, a nonzero chemical potential changes the Fermi surface to a torus~\cite{Rhim:2016dy,Takane:2018fe}.
While the crossover from a 2D to a 3D transport regime was originally predicted for a torus-shaped Fermi surface~\cite{Chen:2019cz}, the conductivity growth with disorder is likely to be affected by moving the chemical potential away from the nodal line.
Furthermore, a BCS-like disorder-driven transition has been predicted for $\mu\neq0$~\cite{Sun:2023ht,Zhu:2023hr} that implies a different scaling of density of states and conductivity above and below a critical disorder strength.
(At $\mu=0$, the critical disorder strength is zero~\cite{Zhu:2023hr}.)
Second, we consider a simplified model with only terms linear in momentum terms contributing to the Hamiltonian.
Terms that are higher order in momentum will introduce more length scales that could provide upper cutoffs for the absence of localization we observe here.
Third, we have neglected surface states throughout this work, which may alter transport properties~\cite{Fu2023}.
Finally, due to the momentum cutoff we used in the numerical simulations, the results are not valid for small values of $L b$.

In addition to this, our results are valid only in the noninteracting regime.
Interaction effects may nontrivially interplay with disorder in nodal line semimetals~\cite{Wang:2017be,Munoz:2020cw}.
This interplay can, unlike in parabolic systems~\cite{Kohn:1961et}, affect transport~\cite{Munoz:2020cw,Syzranov:2017dh}.
Studies specifically addressing how electron interactions influence long-range disorder effects are scarce.
Consequently, it is challenging to predict their interplay, which we leave for future works.

Evading localization is generally linked to topological terms~\cite{Khmelnitskii1983,Pruisken:1984cj,Ostrovsky:2007hc}.
The metal we find for all disorder strengths hints that a topological term might also play a role in nodal line semimetals.
Previous works had found a 3D analog~\cite{Burkov:2018gc,Rui:2018jr} of the $\theta$ term in graphene~\cite{Ostrovsky:2007hc}.
This term might influence transport away from the clean case and could be linked to the absence of localization we found.

Several candidate materials for nodal line semimetals have been identified, including HgCr$_{2}$Se$_{4}$~\cite{Xu2011_ex}, Cu$_{3}$(Pd/Zn)N~\cite{Yu2015_ex,Kim2015_ex}, Mg$_{3}$Bi$_{2}$~\cite{Zhou:2020ex}, and SrIrO$_{3}$~\cite{Chen2015_ex}.
These materials, whether proposed theoretically or confirmed experimentally, present promising opportunities for further experimental investigation.
Transport measurements are crucial for validating our findings as they effectively probe the electronic properties of topological materials, shedding light on the spatial distribution of surface states, the geometric phase of the wave function, and the topology of the Fermi surface.
Transport experiments across a range of impurity concentrations (tunable via doping) could test our theoretical predictions and further investigate the absence of localization across all disorder strengths in these semimetals.

Possible extensions of our work could explore $\mu \neq 0$ and the consequences of the BCS-like transition for transport~\cite{Sun:2023ht,Zhu:2023hr}.
Furthermore, it would be interesting to explore the phase diagram of the potential $V(\mathbf{r}) = (u_0 + u(\mathbf{r}))\Gamma_5$ that opens a gap everywhere in the spectrum.

\begin{acknowledgments}
The authors are grateful to Jens H.\ Bardarson for insightful discussions throughout the project and careful reading of the manuscript.
JB thanks Jun-Won Rhim for helpful discussions, especially in the early stages of this work.
JB is supported by a Leverhulme Early Career Fellowship, and the Newton Trust of the University of Cambridge.

Our simulations used resources at the Cambridge Service for Data Driven Discovery operated by the University of Cambridge Research Computing Service (\href{https://www.csd3.cam.ac.uk}{www.csd3.cam.ac.uk}), provided by Dell EMC and Intel using EPSRC Tier-2 funding via grant EP/T022159/1, and STFC DiRAC funding (\href{https://www.dirac.ac.uk}{www.dirac.ac.uk}).
\end{acknowledgments}

\appendix

\section{\label{sec:mass_term}Nonzero mass term}

\begin{figure}
\includegraphics{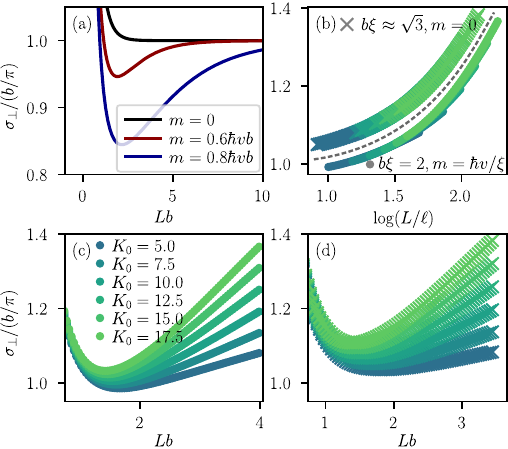}
\caption{(a) Out-of-plane conductivity $\sigma_\perp$ in the clean case for different values of $m$, cf.\ Eq.~\eqref{eq:conductivity_nonzero_mass}.
(b) $\sigma_\perp$ as a function of the rescaled system length $L/\ell$ for different values of $K_0$, where dots denote $b\xi=2$ and $m=\hbar v/\xi$ (with $b_\mathrm{eff}\xi = \sqrt{3}$), and crosses denote $b\xi =1.73 \approx \sqrt{3}$ and $m=0$ (the gray dashed line separates the regimes).
The data does not perfectly collapse when $m\neq 0$.
(c) Raw conductivity data for $m\neq 0$ and (d) for $m=0$ with the same nodal line radius.}
\label{fig:nonzero_mass}
\end{figure}

The focus in the main text is on $m=0$, a fine-tuned point in the phase diagram where the system is effectively time-reversal symmetric.
Here we discuss out-of-plane transport for $m\neq 0$, for both clean and disordered systems with scalar disorder $V(\mathbf{r}) = u(\mathbf{r}) \mathbb{1}$.
In the clean case, i.e., for $u(\mathbf{r})=0$ and for transport parallel to $\mathbf{b} = b \mathbf{e}_z$, we solve the transfer matrix [Eq.~\eqref{eq:transfer_matrix} in the main text], giving the trace
\begin{widetext}
\begin{equation}
 \tr [t^\dagger t] = \frac{1}{\cosh^2 \left[ L \left( b + \sqrt{ k^2 + \tilde{m}^2 } \right) \right]} + \frac{1}{\cosh^2 \left[ L \left( b - \sqrt{ k^2 + \tilde{m}^2 } \right) \right]}
\end{equation}
where $\tilde{m}=m/(\hbar v)$ for a compact notation and $0 \le \tilde{m} < b$ for a gapless system.
In the limit $W \gg L$, the dimensionless conductance is accordingly 
\begin{equation}
 g_\perp = \frac{W^2}{2\pi} \int d k k \tr [t^\dagger t] = \frac{W^2}{L} \frac{b}{\pi} \left( \frac{\log \left[ 2 \left( \cosh ( 2 L \tilde{m})+\cosh (2 L b) \right) \right]}{2 L b} -\frac{ \tilde{m} \sinh (2 L \tilde{m})}{2 b \cosh (L(b-\tilde{m})) \cosh (L(b+\tilde{m}))} \right)
 \label{eq:conductivity_nonzero_mass}
\end{equation}
\end{widetext}
and the conductivity $\sigma_\perp \xrightarrow[L b \gg 1]{} b/\pi$, the same value we obtained in the main text for $m=0$.
We show the $\sigma_\perp$ as a function of $Lb$ for different values of $m$ in Fig.~\ref{fig:nonzero_mass}(a).

The term $m \neq 0$ breaks the emergent time-reversal symmetry discussed in the main text.
To confirm that the behavior we find for disordered nodal line semimetals does not rely on the presence of this emergent time-reversal symmetry, we here consider disordered systems with $m\neq 0$.
In Fig.~\ref{fig:nonzero_mass}(b), we show the rescaled conductivity as a function of $\log (L/\ell)$ for different disorder strengths and for $b\xi=2$, $m/(\hbar v/\xi)=1$ and $b\xi=1.73\approx \sqrt{3}$, $m=0$.
The radius of the nodal line in clean systems is in both cases $b_\mathrm{eff} = \sqrt{b^2 -(m/(\hbar v))^2} \approx \sqrt{3}/\xi$.

Importantly, even when $m\neq 0$, the conductivity increases with both $L$ and $K_0$.
However, the data does not collapse onto the same line for $m\neq 0$, indicating that the 2D scaling does hold only approximately here.
For completeness, we show the raw data in Fig.~\ref{fig:nonzero_mass}(c) and~(d).

\bibliography{nlsm}

\begin{thebibliography}{75}%
\makeatletter
\providecommand \@ifxundefined [1]{%
 \@ifx{#1\undefined}
}%
\providecommand \@ifnum [1]{%
 \ifnum #1\expandafter \@firstoftwo
 \else \expandafter \@secondoftwo
 \fi
}%
\providecommand \@ifx [1]{%
 \ifx #1\expandafter \@firstoftwo
 \else \expandafter \@secondoftwo
 \fi
}%
\providecommand \natexlab [1]{#1}%
\providecommand \enquote  [1]{``#1''}%
\providecommand \bibnamefont  [1]{#1}%
\providecommand \bibfnamefont [1]{#1}%
\providecommand \citenamefont [1]{#1}%
\providecommand \href@noop [0]{\@secondoftwo}%
\providecommand \href [0]{\begingroup \@sanitize@url \@href}%
\providecommand \@href[1]{\@@startlink{#1}\@@href}%
\providecommand \@@href[1]{\endgroup#1\@@endlink}%
\providecommand \@sanitize@url [0]{\catcode `\\12\catcode `\$12\catcode
  `\&12\catcode `\#12\catcode `\^12\catcode `\_12\catcode `\%12\relax}%
\providecommand \@@startlink[1]{}%
\providecommand \@@endlink[0]{}%
\providecommand \url  [0]{\begingroup\@sanitize@url \@url }%
\providecommand \@url [1]{\endgroup\@href {#1}{\urlprefix }}%
\providecommand \urlprefix  [0]{URL }%
\providecommand \Eprint [0]{\href }%
\providecommand \doibase [0]{https://doi.org/}%
\providecommand \selectlanguage [0]{\@gobble}%
\providecommand \bibinfo  [0]{\@secondoftwo}%
\providecommand \bibfield  [0]{\@secondoftwo}%
\providecommand \translation [1]{[#1]}%
\providecommand \BibitemOpen [0]{}%
\providecommand \bibitemStop [0]{}%
\providecommand \bibitemNoStop [0]{.\EOS\space}%
\providecommand \EOS [0]{\spacefactor3000\relax}%
\providecommand \BibitemShut  [1]{\csname bibitem#1\endcsname}%
\let\auto@bib@innerbib\@empty
%</preamble>
\bibitem [{\citenamefont {Abrahams}\ \emph {et~al.}(1979)\citenamefont
  {Abrahams}, \citenamefont {Anderson}, \citenamefont {Licciardello},\ and\
  \citenamefont {Ramakrishnan}}]{Abrahams:1979iv}%
  \BibitemOpen
  \bibfield  {author} {\bibinfo {author} {\bibfnamefont {E.}~\bibnamefont
  {Abrahams}}, \bibinfo {author} {\bibfnamefont {P.~W.}\ \bibnamefont
  {Anderson}}, \bibinfo {author} {\bibfnamefont {D.~C.}\ \bibnamefont
  {Licciardello}},\ and\ \bibinfo {author} {\bibfnamefont {T.~V.}\ \bibnamefont
  {Ramakrishnan}},\ }\bibfield  {title} {\bibinfo {title} {{Scaling Theory of
  Localization: Absence of Quantum Diffusion in Two Dimensions}},\ }\href
  {https://doi.org/10.1103/PhysRevLett.42.673} {\bibfield  {journal} {\bibinfo
  {journal} {Physical Review Letters}\ }\textbf {\bibinfo {volume} {42}},\
  \bibinfo {pages} {673} (\bibinfo {year} {1979})}\BibitemShut {NoStop}%
\bibitem [{\citenamefont {Evers}\ and\ \citenamefont
  {Mirlin}(2008)}]{Evers:2008gi}%
  \BibitemOpen
  \bibfield  {author} {\bibinfo {author} {\bibfnamefont {F.}~\bibnamefont
  {Evers}}\ and\ \bibinfo {author} {\bibfnamefont {A.~D.}\ \bibnamefont
  {Mirlin}},\ }\bibfield  {title} {\bibinfo {title} {{Anderson transitions}},\
  }\href {https://doi.org/10.1103/RevModPhys.80.1355} {\bibfield  {journal}
  {\bibinfo  {journal} {Reviews of Modern Physics}\ }\textbf {\bibinfo {volume}
  {80}},\ \bibinfo {pages} {1355} (\bibinfo {year} {2008})}\BibitemShut
  {NoStop}%
\bibitem [{\citenamefont {Hikami}(1981)}]{Hikami:1981eq}%
  \BibitemOpen
  \bibfield  {author} {\bibinfo {author} {\bibfnamefont {S.}~\bibnamefont
  {Hikami}},\ }\bibfield  {title} {\bibinfo {title} {{Three-loop
  $\beta$-functions of non-linear $\sigma$ models on symmetric spaces}},\
  }\href {https://doi.org/10.1016/0370-2693(81)90989-8} {\bibfield  {journal}
  {\bibinfo  {journal} {Physics Letters B}\ }\textbf {\bibinfo {volume} {98}},\
  \bibinfo {pages} {208} (\bibinfo {year} {1981})}\BibitemShut {NoStop}%
\bibitem [{\citenamefont {Gade}\ and\ \citenamefont
  {Wegner}(1991)}]{Gade:1991ka}%
  \BibitemOpen
  \bibfield  {author} {\bibinfo {author} {\bibfnamefont {R.}~\bibnamefont
  {Gade}}\ and\ \bibinfo {author} {\bibfnamefont {F.}~\bibnamefont {Wegner}},\
  }\bibfield  {title} {\bibinfo {title} {{The $n = 0$ replica limit of $U(n)$
  and $U(n)SO(n)$ models}},\ }\href
  {https://doi.org/10.1016/0550-3213(91)90401-I} {\bibfield  {journal}
  {\bibinfo  {journal} {Nuclear Physics B}\ }\textbf {\bibinfo {volume}
  {360}},\ \bibinfo {pages} {213} (\bibinfo {year} {1991})}\BibitemShut
  {NoStop}%
\bibitem [{\citenamefont {Pruisken}(1984)}]{Pruisken:1984cj}%
  \BibitemOpen
  \bibfield  {author} {\bibinfo {author} {\bibfnamefont {A.}~\bibnamefont
  {Pruisken}},\ }\bibfield  {title} {\bibinfo {title} {{On localization in the
  theory of the quantized hall effect: A two-dimensional realization of the
  $\theta$-vacuum}},\ }\href {https://doi.org/10.1016/0550-3213(84)90101-9}
  {\bibfield  {journal} {\bibinfo  {journal} {Nuclear Physics B}\ }\textbf
  {\bibinfo {volume} {235}},\ \bibinfo {pages} {277} (\bibinfo {year}
  {1984})}\BibitemShut {NoStop}%
\bibitem [{\citenamefont {Ostrovsky}\ \emph {et~al.}(2007)\citenamefont
  {Ostrovsky}, \citenamefont {Gornyi},\ and\ \citenamefont
  {Mirlin}}]{Ostrovsky:2007hc}%
  \BibitemOpen
  \bibfield  {author} {\bibinfo {author} {\bibfnamefont {P.~M.}\ \bibnamefont
  {Ostrovsky}}, \bibinfo {author} {\bibfnamefont {I.~V.}\ \bibnamefont
  {Gornyi}},\ and\ \bibinfo {author} {\bibfnamefont {A.~D.}\ \bibnamefont
  {Mirlin}},\ }\bibfield  {title} {\bibinfo {title} {{Quantum Criticality and
  Minimal Conductivity in Graphene with Long-Range Disorder}},\ }\href
  {https://doi.org/10.1103/PhysRevLett.98.256801} {\bibfield  {journal}
  {\bibinfo  {journal} {Physical Review Letters}\ }\textbf {\bibinfo {volume}
  {98}},\ \bibinfo {pages} {256801} (\bibinfo {year} {2007})}\BibitemShut
  {NoStop}%
\bibitem [{\citenamefont {Bardarson}\ \emph {et~al.}(2007)\citenamefont
  {Bardarson}, \citenamefont {Tworzyd{\l}o}, \citenamefont {Brouwer},\ and\
  \citenamefont {Beenakker}}]{Bardarson:2007iu}%
  \BibitemOpen
  \bibfield  {author} {\bibinfo {author} {\bibfnamefont {J.~H.}\ \bibnamefont
  {Bardarson}}, \bibinfo {author} {\bibfnamefont {J.}~\bibnamefont
  {Tworzyd{\l}o}}, \bibinfo {author} {\bibfnamefont {P.~W.}\ \bibnamefont
  {Brouwer}},\ and\ \bibinfo {author} {\bibfnamefont {C.~W.~J.}\ \bibnamefont
  {Beenakker}},\ }\bibfield  {title} {\bibinfo {title} {{One-Parameter Scaling
  at the Dirac Point in Graphene}},\ }\href
  {https://doi.org/10.1103/PhysRevLett.99.106801} {\bibfield  {journal}
  {\bibinfo  {journal} {Physical Review Letters}\ }\textbf {\bibinfo {volume}
  {99}},\ \bibinfo {pages} {106801} (\bibinfo {year} {2007})}\BibitemShut
  {NoStop}%
\bibitem [{\citenamefont {Nomura}\ \emph {et~al.}(2007)\citenamefont {Nomura},
  \citenamefont {Koshino},\ and\ \citenamefont {Ryu}}]{Nomura:2007jb}%
  \BibitemOpen
  \bibfield  {author} {\bibinfo {author} {\bibfnamefont {K.}~\bibnamefont
  {Nomura}}, \bibinfo {author} {\bibfnamefont {M.}~\bibnamefont {Koshino}},\
  and\ \bibinfo {author} {\bibfnamefont {S.}~\bibnamefont {Ryu}},\ }\bibfield
  {title} {\bibinfo {title} {{Topological Delocalization of Two-Dimensional
  Massless Dirac Fermions}},\ }\href
  {https://doi.org/10.1103/PhysRevLett.99.146806} {\bibfield  {journal}
  {\bibinfo  {journal} {Physical Review Letters}\ }\textbf {\bibinfo {volume}
  {99}},\ \bibinfo {pages} {146806} (\bibinfo {year} {2007})}\BibitemShut
  {NoStop}%
\bibitem [{\citenamefont {Nandkishore}\ \emph {et~al.}(2014)\citenamefont
  {Nandkishore}, \citenamefont {Huse},\ and\ \citenamefont
  {Sondhi}}]{Nandkishore:2014cv}%
  \BibitemOpen
  \bibfield  {author} {\bibinfo {author} {\bibfnamefont {R.}~\bibnamefont
  {Nandkishore}}, \bibinfo {author} {\bibfnamefont {D.~A.}\ \bibnamefont
  {Huse}},\ and\ \bibinfo {author} {\bibfnamefont {S.~L.}\ \bibnamefont
  {Sondhi}},\ }\bibfield  {title} {\bibinfo {title} {{Rare region effects
  dominate weakly disordered three-dimensional Dirac points}},\ }\href
  {https://doi.org/10.1103/PhysRevB.89.245110} {\bibfield  {journal} {\bibinfo
  {journal} {Physical Review B}\ }\textbf {\bibinfo {volume} {89}},\ \bibinfo
  {pages} {245110} (\bibinfo {year} {2014})}\BibitemShut {NoStop}%
\bibitem [{\citenamefont {Pixley}\ and\ \citenamefont
  {Wilson}(2021)}]{Pixley:2021do}%
  \BibitemOpen
  \bibfield  {author} {\bibinfo {author} {\bibfnamefont {J.}~\bibnamefont
  {Pixley}}\ and\ \bibinfo {author} {\bibfnamefont {J.~H.}\ \bibnamefont
  {Wilson}},\ }\bibfield  {title} {\bibinfo {title} {{Rare regions and avoided
  quantum criticality in disordered Weyl semimetals and superconductors}},\
  }\href {https://doi.org/10.1016/j.aop.2021.168455} {\bibfield  {journal}
  {\bibinfo  {journal} {Annals of Physics}\ }\textbf {\bibinfo {volume}
  {435}},\ \bibinfo {pages} {168455} (\bibinfo {year} {2021})}\BibitemShut
  {NoStop}%
\bibitem [{\citenamefont {Goswami}\ and\ \citenamefont
  {Chakravarty}(2011)}]{Goswami:2011hh}%
  \BibitemOpen
  \bibfield  {author} {\bibinfo {author} {\bibfnamefont {P.}~\bibnamefont
  {Goswami}}\ and\ \bibinfo {author} {\bibfnamefont {S.}~\bibnamefont
  {Chakravarty}},\ }\bibfield  {title} {\bibinfo {title} {{Quantum Criticality
  between Topological and Band Insulators in $3+1$ Dimensions}},\ }\href
  {https://doi.org/10.1103/PhysRevLett.107.196803} {\bibfield  {journal}
  {\bibinfo  {journal} {Physical Review Letters}\ }\textbf {\bibinfo {volume}
  {107}},\ \bibinfo {pages} {196803} (\bibinfo {year} {2011})}\BibitemShut
  {NoStop}%
\bibitem [{\citenamefont {Sbierski}\ \emph {et~al.}(2014)\citenamefont
  {Sbierski}, \citenamefont {Pohl}, \citenamefont {Bergholtz},\ and\
  \citenamefont {Brouwer}}]{Sbierski:2014bo}%
  \BibitemOpen
  \bibfield  {author} {\bibinfo {author} {\bibfnamefont {B.}~\bibnamefont
  {Sbierski}}, \bibinfo {author} {\bibfnamefont {G.}~\bibnamefont {Pohl}},
  \bibinfo {author} {\bibfnamefont {E.~J.}\ \bibnamefont {Bergholtz}},\ and\
  \bibinfo {author} {\bibfnamefont {P.~W.}\ \bibnamefont {Brouwer}},\
  }\bibfield  {title} {\bibinfo {title} {{Quantum Transport of Disordered Weyl
  Semimetals at the Nodal Point}},\ }\href
  {https://doi.org/10.1103/PhysRevLett.113.026602} {\bibfield  {journal}
  {\bibinfo  {journal} {Physical Review Letters}\ }\textbf {\bibinfo {volume}
  {113}},\ \bibinfo {pages} {026602} (\bibinfo {year} {2014})}\BibitemShut
  {NoStop}%
\bibitem [{\citenamefont {Syzranov}\ and\ \citenamefont
  {Radzihovsky}(2017)}]{Syzranov:2017hw}%
  \BibitemOpen
  \bibfield  {author} {\bibinfo {author} {\bibfnamefont {S.~V.}\ \bibnamefont
  {Syzranov}}\ and\ \bibinfo {author} {\bibfnamefont {L.}~\bibnamefont
  {Radzihovsky}},\ }\bibfield  {title} {\bibinfo {title} {{High-Dimensional
  Disorder-Driven Phenomena in Weyl Semimetals, Semiconductors and Related
  Systems}},\ }\href {https://doi.org/10.1146/annurev-conmatphys-033117-054037}
  {\bibfield  {journal} {\bibinfo  {journal} {Ann. Rev. Cond. Mat. Phys}\
  }\textbf {\bibinfo {volume} {9}},\ \bibinfo {pages} {35} (\bibinfo {year}
  {2017})}\BibitemShut {NoStop}%
\bibitem [{\citenamefont {Zhu}\ and\ \citenamefont
  {Syzranov}(2023)}]{Zhu:2023hr}%
  \BibitemOpen
  \bibfield  {author} {\bibinfo {author} {\bibfnamefont {S.}~\bibnamefont
  {Zhu}}\ and\ \bibinfo {author} {\bibfnamefont {S.}~\bibnamefont {Syzranov}},\
  }\bibfield  {title} {\bibinfo {title} {{BCS-like disorder-driven
  instabilities and ultraviolet effects in nodal-line semimetals}},\ }\href
  {https://doi.org/10.1016/j.aop.2023.169501} {\bibfield  {journal} {\bibinfo
  {journal} {Annals of Physics}\ }\textbf {\bibinfo {volume} {459}},\ \bibinfo
  {pages} {169501} (\bibinfo {year} {2023})}\BibitemShut {NoStop}%
\bibitem [{\citenamefont {Heikkil{\"{a}}}\ and\ \citenamefont
  {Volovik}(2011)}]{Heikkila:2011hw}%
  \BibitemOpen
  \bibfield  {author} {\bibinfo {author} {\bibfnamefont {T.~T.}\ \bibnamefont
  {Heikkil{\"{a}}}}\ and\ \bibinfo {author} {\bibfnamefont {G.~E.}\
  \bibnamefont {Volovik}},\ }\bibfield  {title} {\bibinfo {title} {{Dimensional
  crossover in topological matter: Evolution of the multiple Dirac point in the
  layered system to the flat band on the surface}},\ }\href
  {https://doi.org/10.1134/S002136401102007X} {\bibfield  {journal} {\bibinfo
  {journal} {JETP Letters}\ }\textbf {\bibinfo {volume} {93}},\ \bibinfo
  {pages} {59} (\bibinfo {year} {2011})}\BibitemShut {NoStop}%
\bibitem [{\citenamefont {B{\'{e}}ri}(2010)}]{Beri:2010dk}%
  \BibitemOpen
  \bibfield  {author} {\bibinfo {author} {\bibfnamefont {B.}~\bibnamefont
  {B{\'{e}}ri}},\ }\bibfield  {title} {\bibinfo {title} {{Topologically stable
  gapless phases of time-reversal-invariant superconductors}},\ }\href
  {https://doi.org/10.1103/PhysRevB.81.134515} {\bibfield  {journal} {\bibinfo
  {journal} {Physical Review B}\ }\textbf {\bibinfo {volume} {81}},\ \bibinfo
  {pages} {134515} (\bibinfo {year} {2010})}\BibitemShut {NoStop}%
\bibitem [{\citenamefont {Burkov}\ \emph {et~al.}(2011)\citenamefont {Burkov},
  \citenamefont {Hook},\ and\ \citenamefont {Balents}}]{Burkov:2011ek}%
  \BibitemOpen
  \bibfield  {author} {\bibinfo {author} {\bibfnamefont {A.~A.}\ \bibnamefont
  {Burkov}}, \bibinfo {author} {\bibfnamefont {M.~D.}\ \bibnamefont {Hook}},\
  and\ \bibinfo {author} {\bibfnamefont {L.}~\bibnamefont {Balents}},\
  }\bibfield  {title} {\bibinfo {title} {{Topological nodal semimetals}},\
  }\href {https://doi.org/10.1103/PhysRevB.84.235126} {\bibfield  {journal}
  {\bibinfo  {journal} {Phys. Rev. B}\ }\textbf {\bibinfo {volume} {84}},\
  \bibinfo {pages} {235126} (\bibinfo {year} {2011})}\BibitemShut {NoStop}%
\bibitem [{\citenamefont {Chiu}\ and\ \citenamefont
  {Schnyder}(2014)}]{Chiu:2014fi}%
  \BibitemOpen
  \bibfield  {author} {\bibinfo {author} {\bibfnamefont {C.-K.}\ \bibnamefont
  {Chiu}}\ and\ \bibinfo {author} {\bibfnamefont {A.~P.}\ \bibnamefont
  {Schnyder}},\ }\bibfield  {title} {\bibinfo {title} {{Classification of
  reflection-symmetry-protected topological semimetals and nodal
  superconductors}},\ }\href {https://doi.org/10.1103/PhysRevB.90.205136}
  {\bibfield  {journal} {\bibinfo  {journal} {Physical Review B}\ }\textbf
  {\bibinfo {volume} {90}},\ \bibinfo {pages} {205136} (\bibinfo {year}
  {2014})}\BibitemShut {NoStop}%
\bibitem [{\citenamefont {Yang}\ \emph {et~al.}(2020)\citenamefont {Yang},
  \citenamefont {Chiu}, \citenamefont {Fang},\ and\ \citenamefont
  {Hu}}]{Yang:2020he}%
  \BibitemOpen
  \bibfield  {author} {\bibinfo {author} {\bibfnamefont {Z.}~\bibnamefont
  {Yang}}, \bibinfo {author} {\bibfnamefont {C.-K.}\ \bibnamefont {Chiu}},
  \bibinfo {author} {\bibfnamefont {C.}~\bibnamefont {Fang}},\ and\ \bibinfo
  {author} {\bibfnamefont {J.}~\bibnamefont {Hu}},\ }\bibfield  {title}
  {\bibinfo {title} {{Jones Polynomial and Knot Transitions in Hermitian and
  non-Hermitian Topological Semimetals}},\ }\href
  {https://doi.org/10.1103/PhysRevLett.124.186402} {\bibfield  {journal}
  {\bibinfo  {journal} {Physical Review Letters}\ }\textbf {\bibinfo {volume}
  {124}},\ \bibinfo {pages} {186402} (\bibinfo {year} {2020})}\BibitemShut
  {NoStop}%
\bibitem [{\citenamefont {Phillips}\ and\ \citenamefont
  {Aji}(2014)}]{Phillips:2014kj}%
  \BibitemOpen
  \bibfield  {author} {\bibinfo {author} {\bibfnamefont {M.}~\bibnamefont
  {Phillips}}\ and\ \bibinfo {author} {\bibfnamefont {V.}~\bibnamefont {Aji}},\
  }\bibfield  {title} {\bibinfo {title} {{Tunable line node semimetals}},\
  }\href {https://doi.org/10.1103/PhysRevB.90.115111} {\bibfield  {journal}
  {\bibinfo  {journal} {Physical Review B}\ }\textbf {\bibinfo {volume} {90}},\
  \bibinfo {pages} {115111} (\bibinfo {year} {2014})}\BibitemShut {NoStop}%
\bibitem [{\citenamefont {Fang}\ \emph {et~al.}(2015)\citenamefont {Fang},
  \citenamefont {Chen}, \citenamefont {Kee},\ and\ \citenamefont
  {Fu}}]{Fang:2015gt}%
  \BibitemOpen
  \bibfield  {author} {\bibinfo {author} {\bibfnamefont {C.}~\bibnamefont
  {Fang}}, \bibinfo {author} {\bibfnamefont {Y.}~\bibnamefont {Chen}}, \bibinfo
  {author} {\bibfnamefont {H.-Y.}\ \bibnamefont {Kee}},\ and\ \bibinfo {author}
  {\bibfnamefont {L.}~\bibnamefont {Fu}},\ }\bibfield  {title} {\bibinfo
  {title} {{Topological nodal line semimetals with and without spin-orbital
  coupling}},\ }\href {https://doi.org/10.1103/PhysRevB.92.081201} {\bibfield
  {journal} {\bibinfo  {journal} {Physical Review B}\ }\textbf {\bibinfo
  {volume} {92}},\ \bibinfo {pages} {081201(R)} (\bibinfo {year}
  {2015})}\BibitemShut {NoStop}%
\bibitem [{\citenamefont {Chan}\ \emph {et~al.}(2016)\citenamefont {Chan},
  \citenamefont {Chiu}, \citenamefont {Chou},\ and\ \citenamefont
  {Schnyder}}]{Chan:2016ho}%
  \BibitemOpen
  \bibfield  {author} {\bibinfo {author} {\bibfnamefont {Y.~H.}\ \bibnamefont
  {Chan}}, \bibinfo {author} {\bibfnamefont {C.~K.}\ \bibnamefont {Chiu}},
  \bibinfo {author} {\bibfnamefont {M.~Y.}\ \bibnamefont {Chou}},\ and\
  \bibinfo {author} {\bibfnamefont {A.~P.}\ \bibnamefont {Schnyder}},\
  }\bibfield  {title} {\bibinfo {title} {{Ca$_3$P$_2$ and other topological
  semimetals with line nodes and drumhead surface states}},\ }\href
  {https://doi.org/10.1103/PhysRevB.93.205132} {\bibfield  {journal} {\bibinfo
  {journal} {Phys. Rev. B}\ }\textbf {\bibinfo {volume} {93}},\ \bibinfo
  {pages} {205132} (\bibinfo {year} {2016})}\BibitemShut {NoStop}%
\bibitem [{\citenamefont {Bzdu{\v{s}}ek}\ and\ \citenamefont
  {Sigrist}(2017)}]{Bzdusek:2017dy}%
  \BibitemOpen
  \bibfield  {author} {\bibinfo {author} {\bibfnamefont {T.}~\bibnamefont
  {Bzdu{\v{s}}ek}}\ and\ \bibinfo {author} {\bibfnamefont {M.}~\bibnamefont
  {Sigrist}},\ }\bibfield  {title} {\bibinfo {title} {{Robust doubly charged
  nodal lines and nodal surfaces in centrosymmetric systems}},\ }\href
  {https://doi.org/10.1103/PhysRevB.96.155105} {\bibfield  {journal} {\bibinfo
  {journal} {Physical Review B}\ }\textbf {\bibinfo {volume} {96}},\ \bibinfo
  {pages} {155105} (\bibinfo {year} {2017})}\BibitemShut {NoStop}%
\bibitem [{\citenamefont {Wang}(2017)}]{Wang:2017dh}%
  \BibitemOpen
  \bibfield  {author} {\bibinfo {author} {\bibfnamefont {J.}~\bibnamefont
  {Wang}},\ }\bibfield  {title} {\bibinfo {title} {{Antiferromagnetic
  topological nodal line semimetals}},\ }\href
  {https://doi.org/10.1103/PhysRevB.96.081107} {\bibfield  {journal} {\bibinfo
  {journal} {Physical Review B}\ }\textbf {\bibinfo {volume} {96}},\ \bibinfo
  {pages} {081107(R)} (\bibinfo {year} {2017})}\BibitemShut {NoStop}%
\bibitem [{\citenamefont {Behrends}\ \emph {et~al.}(2017)\citenamefont
  {Behrends}, \citenamefont {Rhim}, \citenamefont {Liu}, \citenamefont
  {Grushin},\ and\ \citenamefont {Bardarson}}]{Behrends:2017cv}%
  \BibitemOpen
  \bibfield  {author} {\bibinfo {author} {\bibfnamefont {J.}~\bibnamefont
  {Behrends}}, \bibinfo {author} {\bibfnamefont {J.-W.}\ \bibnamefont {Rhim}},
  \bibinfo {author} {\bibfnamefont {S.}~\bibnamefont {Liu}}, \bibinfo {author}
  {\bibfnamefont {A.~G.}\ \bibnamefont {Grushin}},\ and\ \bibinfo {author}
  {\bibfnamefont {J.~H.}\ \bibnamefont {Bardarson}},\ }\bibfield  {title}
  {\bibinfo {title} {{Nodal-line semimetals from Weyl superlattices}},\ }\href
  {https://doi.org/10.1103/PhysRevB.96.245101} {\bibfield  {journal} {\bibinfo
  {journal} {Phys. Rev. B}\ }\textbf {\bibinfo {volume} {96}},\ \bibinfo
  {pages} {245101} (\bibinfo {year} {2017})}\BibitemShut {NoStop}%
\bibitem [{\citenamefont {Bian}\ \emph {et~al.}(2016)\citenamefont {Bian},
  \citenamefont {Chang}, \citenamefont {Sankar}, \citenamefont {Xu},
  \citenamefont {Zheng}, \citenamefont {Neupert}, \citenamefont {Chiu},
  \citenamefont {Huang}, \citenamefont {Chang}, \citenamefont {Belopolski},
  \citenamefont {Sanchez}, \citenamefont {Neupane}, \citenamefont {Alidoust},
  \citenamefont {Liu}, \citenamefont {Wang}, \citenamefont {Lee}, \citenamefont
  {Jeng}, \citenamefont {Zhang}, \citenamefont {Yuan}, \citenamefont {Jia},
  \citenamefont {Bansil}, \citenamefont {Chou}, \citenamefont {Lin},\ and\
  \citenamefont {Hasan}}]{Bian:2016bn}%
  \BibitemOpen
  \bibfield  {author} {\bibinfo {author} {\bibfnamefont {G.}~\bibnamefont
  {Bian}}, \bibinfo {author} {\bibfnamefont {T.-R.}\ \bibnamefont {Chang}},
  \bibinfo {author} {\bibfnamefont {R.}~\bibnamefont {Sankar}}, \bibinfo
  {author} {\bibfnamefont {S.-Y.}\ \bibnamefont {Xu}}, \bibinfo {author}
  {\bibfnamefont {H.}~\bibnamefont {Zheng}}, \bibinfo {author} {\bibfnamefont
  {T.}~\bibnamefont {Neupert}}, \bibinfo {author} {\bibfnamefont {C.-K.}\
  \bibnamefont {Chiu}}, \bibinfo {author} {\bibfnamefont {S.-M.}\ \bibnamefont
  {Huang}}, \bibinfo {author} {\bibfnamefont {G.}~\bibnamefont {Chang}},
  \bibinfo {author} {\bibfnamefont {I.}~\bibnamefont {Belopolski}}, \bibinfo
  {author} {\bibfnamefont {D.~S.}\ \bibnamefont {Sanchez}}, \bibinfo {author}
  {\bibfnamefont {M.}~\bibnamefont {Neupane}}, \bibinfo {author} {\bibfnamefont
  {N.}~\bibnamefont {Alidoust}}, \bibinfo {author} {\bibfnamefont
  {C.}~\bibnamefont {Liu}}, \bibinfo {author} {\bibfnamefont {B.}~\bibnamefont
  {Wang}}, \bibinfo {author} {\bibfnamefont {C.-C.}\ \bibnamefont {Lee}},
  \bibinfo {author} {\bibfnamefont {H.-T.}\ \bibnamefont {Jeng}}, \bibinfo
  {author} {\bibfnamefont {C.}~\bibnamefont {Zhang}}, \bibinfo {author}
  {\bibfnamefont {Z.}~\bibnamefont {Yuan}}, \bibinfo {author} {\bibfnamefont
  {S.}~\bibnamefont {Jia}}, \bibinfo {author} {\bibfnamefont {A.}~\bibnamefont
  {Bansil}}, \bibinfo {author} {\bibfnamefont {F.}~\bibnamefont {Chou}},
  \bibinfo {author} {\bibfnamefont {H.}~\bibnamefont {Lin}},\ and\ \bibinfo
  {author} {\bibfnamefont {M.~Z.}\ \bibnamefont {Hasan}},\ }\bibfield  {title}
  {\bibinfo {title} {{Topological nodal-line fermions in spin-orbit metal
  PbTaSe$_2$}},\ }\href {https://doi.org/10.1038/ncomms10556} {\bibfield
  {journal} {\bibinfo  {journal} {Nature Communications}\ }\textbf {\bibinfo
  {volume} {7}},\ \bibinfo {pages} {10556} (\bibinfo {year}
  {2016})}\BibitemShut {NoStop}%
\bibitem [{\citenamefont {Schoop}\ \emph {et~al.}(2016)\citenamefont {Schoop},
  \citenamefont {Ali}, \citenamefont {Stra{\ss}er}, \citenamefont {Topp},
  \citenamefont {Varykhalov}, \citenamefont {Marchenko}, \citenamefont
  {Duppel}, \citenamefont {Parkin}, \citenamefont {Lotsch},\ and\ \citenamefont
  {Ast}}]{Schoop:2016fv}%
  \BibitemOpen
  \bibfield  {author} {\bibinfo {author} {\bibfnamefont {L.~M.}\ \bibnamefont
  {Schoop}}, \bibinfo {author} {\bibfnamefont {M.~N.}\ \bibnamefont {Ali}},
  \bibinfo {author} {\bibfnamefont {C.}~\bibnamefont {Stra{\ss}er}}, \bibinfo
  {author} {\bibfnamefont {A.}~\bibnamefont {Topp}}, \bibinfo {author}
  {\bibfnamefont {A.}~\bibnamefont {Varykhalov}}, \bibinfo {author}
  {\bibfnamefont {D.}~\bibnamefont {Marchenko}}, \bibinfo {author}
  {\bibfnamefont {V.}~\bibnamefont {Duppel}}, \bibinfo {author} {\bibfnamefont
  {S.~S.~P.}\ \bibnamefont {Parkin}}, \bibinfo {author} {\bibfnamefont {B.~V.}\
  \bibnamefont {Lotsch}},\ and\ \bibinfo {author} {\bibfnamefont {C.~R.}\
  \bibnamefont {Ast}},\ }\bibfield  {title} {\bibinfo {title} {{Dirac cone
  protected by non-symmorphic symmetry and three-dimensional Dirac line node in
  ZrSiS}},\ }\href {https://doi.org/10.1038/ncomms11696} {\bibfield  {journal}
  {\bibinfo  {journal} {Nature Communications}\ }\textbf {\bibinfo {volume}
  {7}},\ \bibinfo {pages} {11696} (\bibinfo {year} {2016})}\BibitemShut
  {NoStop}%
\bibitem [{\citenamefont {Takane}\ \emph {et~al.}(2018)\citenamefont {Takane},
  \citenamefont {Nakayama}, \citenamefont {Souma}, \citenamefont {Wada},
  \citenamefont {Okamoto}, \citenamefont {Takenaka}, \citenamefont {Yamakawa},
  \citenamefont {Yamakage}, \citenamefont {Mitsuhashi}, \citenamefont {Horiba},
  \citenamefont {Kumigashira}, \citenamefont {Takahashi},\ and\ \citenamefont
  {Sato}}]{Takane:2018fe}%
  \BibitemOpen
  \bibfield  {author} {\bibinfo {author} {\bibfnamefont {D.}~\bibnamefont
  {Takane}}, \bibinfo {author} {\bibfnamefont {K.}~\bibnamefont {Nakayama}},
  \bibinfo {author} {\bibfnamefont {S.}~\bibnamefont {Souma}}, \bibinfo
  {author} {\bibfnamefont {T.}~\bibnamefont {Wada}}, \bibinfo {author}
  {\bibfnamefont {Y.}~\bibnamefont {Okamoto}}, \bibinfo {author} {\bibfnamefont
  {K.}~\bibnamefont {Takenaka}}, \bibinfo {author} {\bibfnamefont
  {Y.}~\bibnamefont {Yamakawa}}, \bibinfo {author} {\bibfnamefont
  {A.}~\bibnamefont {Yamakage}}, \bibinfo {author} {\bibfnamefont
  {T.}~\bibnamefont {Mitsuhashi}}, \bibinfo {author} {\bibfnamefont
  {K.}~\bibnamefont {Horiba}}, \bibinfo {author} {\bibfnamefont
  {H.}~\bibnamefont {Kumigashira}}, \bibinfo {author} {\bibfnamefont
  {T.}~\bibnamefont {Takahashi}},\ and\ \bibinfo {author} {\bibfnamefont
  {T.}~\bibnamefont {Sato}},\ }\bibfield  {title} {\bibinfo {title}
  {{Observation of Dirac-like energy band and ring-torus Fermi surface
  associated with the nodal line in topological insulator CaAgAs}},\ }\href
  {https://doi.org/10.1038/s41535-017-0074-z} {\bibfield  {journal} {\bibinfo
  {journal} {npj Quantum Materials}\ }\textbf {\bibinfo {volume} {3}},\
  \bibinfo {pages} {1} (\bibinfo {year} {2018})}\BibitemShut {NoStop}%
\bibitem [{\citenamefont {Chang}\ \emph {et~al.}(2019)\citenamefont {Chang},
  \citenamefont {Pletikosic}, \citenamefont {Kong}, \citenamefont {Bian},
  \citenamefont {Huang}, \citenamefont {Denlinger}, \citenamefont {Kushwaha},
  \citenamefont {Sinkovic}, \citenamefont {Jeng}, \citenamefont {Valla},
  \citenamefont {Xie},\ and\ \citenamefont {Cava}}]{Chang:2019kz}%
  \BibitemOpen
  \bibfield  {author} {\bibinfo {author} {\bibfnamefont {T.-R.}\ \bibnamefont
  {Chang}}, \bibinfo {author} {\bibfnamefont {I.}~\bibnamefont {Pletikosic}},
  \bibinfo {author} {\bibfnamefont {T.}~\bibnamefont {Kong}}, \bibinfo {author}
  {\bibfnamefont {G.}~\bibnamefont {Bian}}, \bibinfo {author} {\bibfnamefont
  {A.}~\bibnamefont {Huang}}, \bibinfo {author} {\bibfnamefont
  {J.}~\bibnamefont {Denlinger}}, \bibinfo {author} {\bibfnamefont {S.~K.}\
  \bibnamefont {Kushwaha}}, \bibinfo {author} {\bibfnamefont {B.}~\bibnamefont
  {Sinkovic}}, \bibinfo {author} {\bibfnamefont {H.-T.}\ \bibnamefont {Jeng}},
  \bibinfo {author} {\bibfnamefont {T.}~\bibnamefont {Valla}}, \bibinfo
  {author} {\bibfnamefont {W.}~\bibnamefont {Xie}},\ and\ \bibinfo {author}
  {\bibfnamefont {R.~J.}\ \bibnamefont {Cava}},\ }\bibfield  {title} {\bibinfo
  {title} {{Realization of a Type‐II Nodal‐Line Semimetal in
  Mg$_3$Bi$_2$}},\ }\href {https://doi.org/10.1002/advs.201800897} {\bibfield
  {journal} {\bibinfo  {journal} {Advanced Science}\ }\textbf {\bibinfo
  {volume} {6}},\ \bibinfo {pages} {1800897} (\bibinfo {year}
  {2019})}\BibitemShut {NoStop}%
\bibitem [{\citenamefont {Fu}\ \emph {et~al.}(2019)\citenamefont {Fu},
  \citenamefont {Yi}, \citenamefont {Zhang}, \citenamefont {Caputo},
  \citenamefont {Ma}, \citenamefont {Gao}, \citenamefont {Lv}, \citenamefont
  {Kong}, \citenamefont {Huang}, \citenamefont {Richard}, \citenamefont {Shi},
  \citenamefont {Strocov}, \citenamefont {Fang}, \citenamefont {Weng},
  \citenamefont {Shi}, \citenamefont {Qian},\ and\ \citenamefont
  {Ding}}]{Fu:2019gh}%
  \BibitemOpen
  \bibfield  {author} {\bibinfo {author} {\bibfnamefont {B.-B.}\ \bibnamefont
  {Fu}}, \bibinfo {author} {\bibfnamefont {C.-J.}\ \bibnamefont {Yi}}, \bibinfo
  {author} {\bibfnamefont {T.-T.}\ \bibnamefont {Zhang}}, \bibinfo {author}
  {\bibfnamefont {M.}~\bibnamefont {Caputo}}, \bibinfo {author} {\bibfnamefont
  {J.-Z.}\ \bibnamefont {Ma}}, \bibinfo {author} {\bibfnamefont
  {X.}~\bibnamefont {Gao}}, \bibinfo {author} {\bibfnamefont {B.~Q.}\
  \bibnamefont {Lv}}, \bibinfo {author} {\bibfnamefont {L.-Y.}\ \bibnamefont
  {Kong}}, \bibinfo {author} {\bibfnamefont {Y.-B.}\ \bibnamefont {Huang}},
  \bibinfo {author} {\bibfnamefont {P.}~\bibnamefont {Richard}}, \bibinfo
  {author} {\bibfnamefont {M.}~\bibnamefont {Shi}}, \bibinfo {author}
  {\bibfnamefont {V.~N.}\ \bibnamefont {Strocov}}, \bibinfo {author}
  {\bibfnamefont {C.}~\bibnamefont {Fang}}, \bibinfo {author} {\bibfnamefont
  {H.-M.}\ \bibnamefont {Weng}}, \bibinfo {author} {\bibfnamefont {Y.-G.}\
  \bibnamefont {Shi}}, \bibinfo {author} {\bibfnamefont {T.}~\bibnamefont
  {Qian}},\ and\ \bibinfo {author} {\bibfnamefont {H.}~\bibnamefont {Ding}},\
  }\bibfield  {title} {\bibinfo {title} {{Dirac nodal surfaces and nodal lines
  in ZrSiS}},\ }\href {https://doi.org/10.1126/sciadv.aau6459} {\bibfield
  {journal} {\bibinfo  {journal} {Science Advances}\ }\textbf {\bibinfo
  {volume} {5}},\ \bibinfo {pages} {1} (\bibinfo {year} {2019})}\BibitemShut
  {NoStop}%
\bibitem [{\citenamefont {Lv}\ \emph {et~al.}(2021)\citenamefont {Lv},
  \citenamefont {Qian},\ and\ \citenamefont {Ding}}]{Lv:2021er}%
  \BibitemOpen
  \bibfield  {author} {\bibinfo {author} {\bibfnamefont {B.~Q.}\ \bibnamefont
  {Lv}}, \bibinfo {author} {\bibfnamefont {T.}~\bibnamefont {Qian}},\ and\
  \bibinfo {author} {\bibfnamefont {H.}~\bibnamefont {Ding}},\ }\bibfield
  {title} {\bibinfo {title} {{Experimental perspective on three-dimensional
  topological semimetals}},\ }\href
  {https://doi.org/10.1103/RevModPhys.93.025002} {\bibfield  {journal}
  {\bibinfo  {journal} {Reviews of Modern Physics}\ }\textbf {\bibinfo {volume}
  {93}},\ \bibinfo {pages} {025002} (\bibinfo {year} {2021})}\BibitemShut
  {NoStop}%
\bibitem [{\citenamefont {Deng}\ \emph {et~al.}(2019)\citenamefont {Deng},
  \citenamefont {Lu}, \citenamefont {Li}, \citenamefont {Huang}, \citenamefont
  {Yan}, \citenamefont {Ma},\ and\ \citenamefont {Liu}}]{Deng:2019kj}%
  \BibitemOpen
  \bibfield  {author} {\bibinfo {author} {\bibfnamefont {W.}~\bibnamefont
  {Deng}}, \bibinfo {author} {\bibfnamefont {J.}~\bibnamefont {Lu}}, \bibinfo
  {author} {\bibfnamefont {F.}~\bibnamefont {Li}}, \bibinfo {author}
  {\bibfnamefont {X.}~\bibnamefont {Huang}}, \bibinfo {author} {\bibfnamefont
  {M.}~\bibnamefont {Yan}}, \bibinfo {author} {\bibfnamefont {J.}~\bibnamefont
  {Ma}},\ and\ \bibinfo {author} {\bibfnamefont {Z.}~\bibnamefont {Liu}},\
  }\bibfield  {title} {\bibinfo {title} {{Nodal rings and drumhead surface
  states in phononic crystals}},\ }\href
  {https://doi.org/10.1038/s41467-019-09820-8} {\bibfield  {journal} {\bibinfo
  {journal} {Nature Communications}\ }\textbf {\bibinfo {volume} {10}},\
  \bibinfo {pages} {1769} (\bibinfo {year} {2019})}\BibitemShut {NoStop}%
\bibitem [{\citenamefont {Song}\ \emph {et~al.}(2019)\citenamefont {Song},
  \citenamefont {He}, \citenamefont {Niu}, \citenamefont {Zhang}, \citenamefont
  {Ren}, \citenamefont {Liu},\ and\ \citenamefont {Jo}}]{Song:2019gl}%
  \BibitemOpen
  \bibfield  {author} {\bibinfo {author} {\bibfnamefont {B.}~\bibnamefont
  {Song}}, \bibinfo {author} {\bibfnamefont {C.}~\bibnamefont {He}}, \bibinfo
  {author} {\bibfnamefont {S.}~\bibnamefont {Niu}}, \bibinfo {author}
  {\bibfnamefont {L.}~\bibnamefont {Zhang}}, \bibinfo {author} {\bibfnamefont
  {Z.}~\bibnamefont {Ren}}, \bibinfo {author} {\bibfnamefont {X.-J.}\
  \bibnamefont {Liu}},\ and\ \bibinfo {author} {\bibfnamefont {G.-B.}\
  \bibnamefont {Jo}},\ }\bibfield  {title} {\bibinfo {title} {{Observation of
  nodal-line semimetal with ultracold fermions in an optical lattice}},\ }\href
  {https://doi.org/10.1038/s41567-019-0564-y} {\bibfield  {journal} {\bibinfo
  {journal} {Nature Physics}\ }\textbf {\bibinfo {volume} {15}},\ \bibinfo
  {pages} {911} (\bibinfo {year} {2019})}\BibitemShut {NoStop}%
\bibitem [{\citenamefont {Syzranov}\ and\ \citenamefont
  {Skinner}(2017)}]{Syzranov:2017dh}%
  \BibitemOpen
  \bibfield  {author} {\bibinfo {author} {\bibfnamefont {S.~V.}\ \bibnamefont
  {Syzranov}}\ and\ \bibinfo {author} {\bibfnamefont {B.}~\bibnamefont
  {Skinner}},\ }\bibfield  {title} {\bibinfo {title} {{Electron transport in
  nodal-line semimetals}},\ }\href {https://doi.org/10.1103/PhysRevB.96.161105}
  {\bibfield  {journal} {\bibinfo  {journal} {Physical Review B}\ }\textbf
  {\bibinfo {volume} {96}},\ \bibinfo {pages} {161105(R)} (\bibinfo {year}
  {2017})}\BibitemShut {NoStop}%
\bibitem [{\citenamefont {Chen}\ \emph {et~al.}(2019)\citenamefont {Chen},
  \citenamefont {Lu},\ and\ \citenamefont {Zilberberg}}]{Chen:2019cz}%
  \BibitemOpen
  \bibfield  {author} {\bibinfo {author} {\bibfnamefont {W.}~\bibnamefont
  {Chen}}, \bibinfo {author} {\bibfnamefont {H.-Z.}\ \bibnamefont {Lu}},\ and\
  \bibinfo {author} {\bibfnamefont {O.}~\bibnamefont {Zilberberg}},\ }\bibfield
   {title} {\bibinfo {title} {{Weak Localization and Antilocalization in
  Nodal-Line Semimetals: Dimensionality and Topological Effects}},\ }\href
  {https://doi.org/10.1103/PhysRevLett.122.196603} {\bibfield  {journal}
  {\bibinfo  {journal} {Physical Review Letters}\ }\textbf {\bibinfo {volume}
  {122}},\ \bibinfo {pages} {196603} (\bibinfo {year} {2019})}\BibitemShut
  {NoStop}%
\bibitem [{\citenamefont {Yang}\ \emph {et~al.}(2022)\citenamefont {Yang},
  \citenamefont {Luo},\ and\ \citenamefont {Chen}}]{Yang:2022ct}%
  \BibitemOpen
  \bibfield  {author} {\bibinfo {author} {\bibfnamefont {M.-X.}\ \bibnamefont
  {Yang}}, \bibinfo {author} {\bibfnamefont {W.}~\bibnamefont {Luo}},\ and\
  \bibinfo {author} {\bibfnamefont {W.}~\bibnamefont {Chen}},\ }\bibfield
  {title} {\bibinfo {title} {{Quantum transport in topological nodal-line
  semimetals}},\ }\href {https://doi.org/10.1080/23746149.2022.2065216}
  {\bibfield  {journal} {\bibinfo  {journal} {Advances in Physics: X}\ }\textbf
  {\bibinfo {volume} {7}},\ \bibinfo {pages} {2065216} (\bibinfo {year}
  {2022})}\BibitemShut {NoStop}%
\bibitem [{\citenamefont {Gon\ifmmode~\mbox{\c{c}}\else \c{c}\fi{}alves}\ \emph
  {et~al.}(2020)\citenamefont {Gon\ifmmode~\mbox{\c{c}}\else \c{c}\fi{}alves},
  \citenamefont {Ribeiro}, \citenamefont {Castro},\ and\ \citenamefont
  {Ara\'ujo}}]{Goncalves:2020ci}%
  \BibitemOpen
  \bibfield  {author} {\bibinfo {author} {\bibfnamefont {M.}~\bibnamefont
  {Gon\ifmmode~\mbox{\c{c}}\else \c{c}\fi{}alves}}, \bibinfo {author}
  {\bibfnamefont {P.}~\bibnamefont {Ribeiro}}, \bibinfo {author} {\bibfnamefont
  {E.~V.}\ \bibnamefont {Castro}},\ and\ \bibinfo {author} {\bibfnamefont
  {M.~A.~N.}\ \bibnamefont {Ara\'ujo}},\ }\bibfield  {title} {\bibinfo {title}
  {{Disorder-Driven Multifractality Transition in Weyl Nodal Loops}},\ }\href
  {https://doi.org/10.1103/PhysRevLett.124.136405} {\bibfield  {journal}
  {\bibinfo  {journal} {Physical Review Letters}\ }\textbf {\bibinfo {volume}
  {124}},\ \bibinfo {pages} {136405} (\bibinfo {year} {2020})}\BibitemShut
  {NoStop}%
\bibitem [{\citenamefont {Mukherjee}\ and\ \citenamefont
  {Carbotte}(2017)}]{Mukherjee:2017fe}%
  \BibitemOpen
  \bibfield  {author} {\bibinfo {author} {\bibfnamefont {S.~P.}\ \bibnamefont
  {Mukherjee}}\ and\ \bibinfo {author} {\bibfnamefont {J.~P.}\ \bibnamefont
  {Carbotte}},\ }\bibfield  {title} {\bibinfo {title} {{Transport and optics at
  the node in a nodal loop semimetal}},\ }\href
  {https://doi.org/10.1103/PhysRevB.95.214203} {\bibfield  {journal} {\bibinfo
  {journal} {Physical Review B}\ }\textbf {\bibinfo {volume} {95}},\ \bibinfo
  {pages} {214203} (\bibinfo {year} {2017})}\BibitemShut {NoStop}%
\bibitem [{\citenamefont {An}\ \emph {et~al.}(2019)\citenamefont {An},
  \citenamefont {Zhu}, \citenamefont {Gao}, \citenamefont {Wu}, \citenamefont
  {Ning},\ and\ \citenamefont {Tian}}]{An:2019ko}%
  \BibitemOpen
  \bibfield  {author} {\bibinfo {author} {\bibfnamefont {L.}~\bibnamefont
  {An}}, \bibinfo {author} {\bibfnamefont {X.}~\bibnamefont {Zhu}}, \bibinfo
  {author} {\bibfnamefont {W.}~\bibnamefont {Gao}}, \bibinfo {author}
  {\bibfnamefont {M.}~\bibnamefont {Wu}}, \bibinfo {author} {\bibfnamefont
  {W.}~\bibnamefont {Ning}},\ and\ \bibinfo {author} {\bibfnamefont
  {M.}~\bibnamefont {Tian}},\ }\bibfield  {title} {\bibinfo {title} {{Chiral
  anomaly and nontrivial Berry phase in the topological nodal-line semimetal
  SrAs$_3$}},\ }\href {https://doi.org/10.1103/PhysRevB.99.045143} {\bibfield
  {journal} {\bibinfo  {journal} {Physical Review B}\ }\textbf {\bibinfo
  {volume} {99}},\ \bibinfo {pages} {045143} (\bibinfo {year}
  {2019})}\BibitemShut {NoStop}%
\bibitem [{\citenamefont {Laha}\ \emph {et~al.}(2019)\citenamefont {Laha},
  \citenamefont {Malick}, \citenamefont {Singha}, \citenamefont {Mandal},
  \citenamefont {Rambabu}, \citenamefont {Kanchana},\ and\ \citenamefont
  {Hossain}}]{Laha:2019ei}%
  \BibitemOpen
  \bibfield  {author} {\bibinfo {author} {\bibfnamefont {A.}~\bibnamefont
  {Laha}}, \bibinfo {author} {\bibfnamefont {S.}~\bibnamefont {Malick}},
  \bibinfo {author} {\bibfnamefont {R.}~\bibnamefont {Singha}}, \bibinfo
  {author} {\bibfnamefont {P.}~\bibnamefont {Mandal}}, \bibinfo {author}
  {\bibfnamefont {P.}~\bibnamefont {Rambabu}}, \bibinfo {author} {\bibfnamefont
  {V.}~\bibnamefont {Kanchana}},\ and\ \bibinfo {author} {\bibfnamefont
  {Z.}~\bibnamefont {Hossain}},\ }\bibfield  {title} {\bibinfo {title}
  {{Magnetotransport properties of the correlated topological nodal-line
  semimetal YbCdGe}},\ }\href {https://doi.org/10.1103/PhysRevB.99.241102}
  {\bibfield  {journal} {\bibinfo  {journal} {Physical Review B}\ }\textbf
  {\bibinfo {volume} {99}},\ \bibinfo {pages} {241102(R)} (\bibinfo {year}
  {2019})}\BibitemShut {NoStop}%
\bibitem [{\citenamefont {Zhou}\ \emph {et~al.}(2020)\citenamefont {Zhou},
  \citenamefont {Tong}, \citenamefont {Xie}, \citenamefont {Yu}, \citenamefont
  {Zhu}, \citenamefont {Wang},\ and\ \citenamefont {Jiang}}]{Zhou:2020ex}%
  \BibitemOpen
  \bibfield  {author} {\bibinfo {author} {\bibfnamefont {T.}~\bibnamefont
  {Zhou}}, \bibinfo {author} {\bibfnamefont {M.}~\bibnamefont {Tong}}, \bibinfo
  {author} {\bibfnamefont {X.}~\bibnamefont {Xie}}, \bibinfo {author}
  {\bibfnamefont {Y.}~\bibnamefont {Yu}}, \bibinfo {author} {\bibfnamefont
  {X.}~\bibnamefont {Zhu}}, \bibinfo {author} {\bibfnamefont {Z.-Y.}\
  \bibnamefont {Wang}},\ and\ \bibinfo {author} {\bibfnamefont
  {T.}~\bibnamefont {Jiang}},\ }\bibfield  {title} {\bibinfo {title} {{Quantum
  Transport Signatures of a Close Candidate for a Type II Nodal-Line
  Semimetal}},\ }\href {https://doi.org/10.1021/acs.jpclett.0c01726} {\bibfield
   {journal} {\bibinfo  {journal} {The Journal of Physical Chemistry Letters}\
  }\textbf {\bibinfo {volume} {11}},\ \bibinfo {pages} {6475} (\bibinfo {year}
  {2020})}\BibitemShut {NoStop}%
\bibitem [{\citenamefont {Cheianov}\ and\ \citenamefont
  {Fal'ko}(2006)}]{Cheianov:2006hl}%
  \BibitemOpen
  \bibfield  {author} {\bibinfo {author} {\bibfnamefont {V.~V.}\ \bibnamefont
  {Cheianov}}\ and\ \bibinfo {author} {\bibfnamefont {V.~I.}\ \bibnamefont
  {Fal'ko}},\ }\bibfield  {title} {\bibinfo {title} {{Selective transmission of
  Dirac electrons and ballistic magnetoresistance of $n$-$p$ junctions in
  graphene}},\ }\href {https://doi.org/10.1103/PhysRevB.74.041403} {\bibfield
  {journal} {\bibinfo  {journal} {Physical Review B}\ }\textbf {\bibinfo
  {volume} {74}},\ \bibinfo {pages} {041403(R)} (\bibinfo {year}
  {2006})}\BibitemShut {NoStop}%
\bibitem [{\citenamefont {Titov}(2007)}]{Titov:2007ev}%
  \BibitemOpen
  \bibfield  {author} {\bibinfo {author} {\bibfnamefont {M.}~\bibnamefont
  {Titov}},\ }\bibfield  {title} {\bibinfo {title} {{Impurity-assisted
  tunneling in graphene}},\ }\href {https://doi.org/10.1209/0295-5075/79/17004}
  {\bibfield  {journal} {\bibinfo  {journal} {Europhysics Letters (EPL)}\
  }\textbf {\bibinfo {volume} {79}},\ \bibinfo {pages} {17004} (\bibinfo {year}
  {2007})}\BibitemShut {NoStop}%
\bibitem [{\citenamefont {Mello}\ \emph {et~al.}(1988)\citenamefont {Mello},
  \citenamefont {Pereyra},\ and\ \citenamefont {Kumar}}]{Mello:1988cj}%
  \BibitemOpen
  \bibfield  {author} {\bibinfo {author} {\bibfnamefont {P.~A.}\ \bibnamefont
  {Mello}}, \bibinfo {author} {\bibfnamefont {P.}~\bibnamefont {Pereyra}},\
  and\ \bibinfo {author} {\bibfnamefont {N.}~\bibnamefont {Kumar}},\ }\bibfield
   {title} {\bibinfo {title} {{Macroscopic approach to multichannel disordered
  conductors}},\ }\href {https://doi.org/10.1016/0003-4916(88)90169-8}
  {\bibfield  {journal} {\bibinfo  {journal} {Annals of Physics}\ }\textbf
  {\bibinfo {volume} {181}},\ \bibinfo {pages} {290} (\bibinfo {year}
  {1988})}\BibitemShut {NoStop}%
\bibitem [{\citenamefont {Tamura}\ and\ \citenamefont
  {Ando}(1991)}]{Tamura:1991ki}%
  \BibitemOpen
  \bibfield  {author} {\bibinfo {author} {\bibfnamefont {H.}~\bibnamefont
  {Tamura}}\ and\ \bibinfo {author} {\bibfnamefont {T.}~\bibnamefont {Ando}},\
  }\bibfield  {title} {\bibinfo {title} {{Conductance fluctuations in quantum
  wires}},\ }\href {https://doi.org/10.1103/PhysRevB.44.1792} {\bibfield
  {journal} {\bibinfo  {journal} {Physical Review B}\ }\textbf {\bibinfo
  {volume} {44}},\ \bibinfo {pages} {1792} (\bibinfo {year}
  {1991})}\BibitemShut {NoStop}%
\bibitem [{\citenamefont {Beenakker}(1997)}]{Beenakker:1997gz}%
  \BibitemOpen
  \bibfield  {author} {\bibinfo {author} {\bibfnamefont {C.~W.~J.}\
  \bibnamefont {Beenakker}},\ }\bibfield  {title} {\bibinfo {title}
  {{Random-matrix theory of quantum transport}},\ }\href
  {https://doi.org/10.1103/RevModPhys.69.731} {\bibfield  {journal} {\bibinfo
  {journal} {Reviews of Modern Physics}\ }\textbf {\bibinfo {volume} {69}},\
  \bibinfo {pages} {731} (\bibinfo {year} {1997})}\BibitemShut {NoStop}%
\bibitem [{\citenamefont {Prada}\ \emph {et~al.}(2007)\citenamefont {Prada},
  \citenamefont {San-Jose}, \citenamefont {Wunsch},\ and\ \citenamefont
  {Guinea}}]{Prada:2007im}%
  \BibitemOpen
  \bibfield  {author} {\bibinfo {author} {\bibfnamefont {E.}~\bibnamefont
  {Prada}}, \bibinfo {author} {\bibfnamefont {P.}~\bibnamefont {San-Jose}},
  \bibinfo {author} {\bibfnamefont {B.}~\bibnamefont {Wunsch}},\ and\ \bibinfo
  {author} {\bibfnamefont {F.}~\bibnamefont {Guinea}},\ }\bibfield  {title}
  {\bibinfo {title} {{Pseudodiffusive magnetotransport in graphene}},\ }\href
  {https://doi.org/10.1103/PhysRevB.75.113407} {\bibfield  {journal} {\bibinfo
  {journal} {Physical Review B}\ }\textbf {\bibinfo {volume} {75}},\ \bibinfo
  {pages} {113407} (\bibinfo {year} {2007})}\BibitemShut {NoStop}%
\bibitem [{\citenamefont {Novoselov}\ \emph {et~al.}(2005)\citenamefont
  {Novoselov}, \citenamefont {Geim}, \citenamefont {Morozov}, \citenamefont
  {Jiang}, \citenamefont {Katsnelson}, \citenamefont {Grigorieva},
  \citenamefont {Dubonos},\ and\ \citenamefont {Firsov}}]{Novoselov:2005es}%
  \BibitemOpen
  \bibfield  {author} {\bibinfo {author} {\bibfnamefont {K.~S.}\ \bibnamefont
  {Novoselov}}, \bibinfo {author} {\bibfnamefont {A.~K.}\ \bibnamefont {Geim}},
  \bibinfo {author} {\bibfnamefont {S.~V.}\ \bibnamefont {Morozov}}, \bibinfo
  {author} {\bibfnamefont {D.}~\bibnamefont {Jiang}}, \bibinfo {author}
  {\bibfnamefont {M.~I.}\ \bibnamefont {Katsnelson}}, \bibinfo {author}
  {\bibfnamefont {I.~V.}\ \bibnamefont {Grigorieva}}, \bibinfo {author}
  {\bibfnamefont {S.~V.}\ \bibnamefont {Dubonos}},\ and\ \bibinfo {author}
  {\bibfnamefont {A.~A.}\ \bibnamefont {Firsov}},\ }\bibfield  {title}
  {\bibinfo {title} {{Two-dimensional gas of massless Dirac fermions in
  graphene}},\ }\href {https://doi.org/10.1038/nature04233} {\bibfield
  {journal} {\bibinfo  {journal} {Nature}\ }\textbf {\bibinfo {volume} {438}},\
  \bibinfo {pages} {197} (\bibinfo {year} {2005})}\BibitemShut {NoStop}%
\bibitem [{\citenamefont {Xiong}\ \emph {et~al.}(2015)\citenamefont {Xiong},
  \citenamefont {Kushwaha}, \citenamefont {Liang}, \citenamefont {Krizan},
  \citenamefont {Hirschberger}, \citenamefont {Wang}, \citenamefont {Cava},\
  and\ \citenamefont {Ong}}]{Xiong:2015kl}%
  \BibitemOpen
  \bibfield  {author} {\bibinfo {author} {\bibfnamefont {J.}~\bibnamefont
  {Xiong}}, \bibinfo {author} {\bibfnamefont {S.~K.}\ \bibnamefont {Kushwaha}},
  \bibinfo {author} {\bibfnamefont {T.}~\bibnamefont {Liang}}, \bibinfo
  {author} {\bibfnamefont {J.~W.}\ \bibnamefont {Krizan}}, \bibinfo {author}
  {\bibfnamefont {M.}~\bibnamefont {Hirschberger}}, \bibinfo {author}
  {\bibfnamefont {W.}~\bibnamefont {Wang}}, \bibinfo {author} {\bibfnamefont
  {R.~J.}\ \bibnamefont {Cava}},\ and\ \bibinfo {author} {\bibfnamefont
  {N.~P.}\ \bibnamefont {Ong}},\ }\bibfield  {title} {\bibinfo {title}
  {{Evidence for the chiral anomaly in the Dirac semimetal Na$_3$Bi}},\ }\href
  {https://doi.org/10.1126/science.aac6089} {\bibfield  {journal} {\bibinfo
  {journal} {Science}\ }\textbf {\bibinfo {volume} {350}},\ \bibinfo {pages}
  {413} (\bibinfo {year} {2015})}\BibitemShut {NoStop}%
\bibitem [{\citenamefont {Gooth}\ \emph {et~al.}(2017)\citenamefont {Gooth},
  \citenamefont {Niemann}, \citenamefont {Meng}, \citenamefont {Grushin},
  \citenamefont {Landsteiner}, \citenamefont {Gotsmann}, \citenamefont
  {Menges}, \citenamefont {Schmidt}, \citenamefont {Shekhar}, \citenamefont
  {S{\"{u}}{\ss}}, \citenamefont {H{\"{u}}hne}, \citenamefont {Rellinghaus},
  \citenamefont {Felser}, \citenamefont {Yan},\ and\ \citenamefont
  {Nielsch}}]{Gooth:2017bn}%
  \BibitemOpen
  \bibfield  {author} {\bibinfo {author} {\bibfnamefont {J.}~\bibnamefont
  {Gooth}}, \bibinfo {author} {\bibfnamefont {A.~C.}\ \bibnamefont {Niemann}},
  \bibinfo {author} {\bibfnamefont {T.}~\bibnamefont {Meng}}, \bibinfo {author}
  {\bibfnamefont {A.~G.}\ \bibnamefont {Grushin}}, \bibinfo {author}
  {\bibfnamefont {K.}~\bibnamefont {Landsteiner}}, \bibinfo {author}
  {\bibfnamefont {B.}~\bibnamefont {Gotsmann}}, \bibinfo {author}
  {\bibfnamefont {F.}~\bibnamefont {Menges}}, \bibinfo {author} {\bibfnamefont
  {M.}~\bibnamefont {Schmidt}}, \bibinfo {author} {\bibfnamefont
  {C.}~\bibnamefont {Shekhar}}, \bibinfo {author} {\bibfnamefont
  {V.}~\bibnamefont {S{\"{u}}{\ss}}}, \bibinfo {author} {\bibfnamefont
  {R.}~\bibnamefont {H{\"{u}}hne}}, \bibinfo {author} {\bibfnamefont
  {B.}~\bibnamefont {Rellinghaus}}, \bibinfo {author} {\bibfnamefont
  {C.}~\bibnamefont {Felser}}, \bibinfo {author} {\bibfnamefont
  {B.}~\bibnamefont {Yan}},\ and\ \bibinfo {author} {\bibfnamefont
  {K.}~\bibnamefont {Nielsch}},\ }\bibfield  {title} {\bibinfo {title}
  {{Experimental signatures of the mixed axial–gravitational anomaly in the
  Weyl semimetal NbP}},\ }\href {https://doi.org/10.1038/nature23005}
  {\bibfield  {journal} {\bibinfo  {journal} {Nature}\ }\textbf {\bibinfo
  {volume} {547}},\ \bibinfo {pages} {324} (\bibinfo {year}
  {2017})}\BibitemShut {NoStop}%
\bibitem [{\citenamefont {Xu}\ \emph {et~al.}(2011)\citenamefont {Xu},
  \citenamefont {Weng}, \citenamefont {Wang}, \citenamefont {Dai},\ and\
  \citenamefont {Fang}}]{Xu2011_ex}%
  \BibitemOpen
  \bibfield  {author} {\bibinfo {author} {\bibfnamefont {G.}~\bibnamefont
  {Xu}}, \bibinfo {author} {\bibfnamefont {H.}~\bibnamefont {Weng}}, \bibinfo
  {author} {\bibfnamefont {Z.}~\bibnamefont {Wang}}, \bibinfo {author}
  {\bibfnamefont {X.}~\bibnamefont {Dai}},\ and\ \bibinfo {author}
  {\bibfnamefont {Z.}~\bibnamefont {Fang}},\ }\bibfield  {title} {\bibinfo
  {title} {{Chern Semimetal and the Quantized Anomalous Hall Effect in
  ${\mathrm{HgCr}}_{2}{\mathrm{Se}}_{4}$}},\ }\href
  {https://doi.org/10.1103/PhysRevLett.107.186806} {\bibfield  {journal}
  {\bibinfo  {journal} {Physical Review Letters}\ }\textbf {\bibinfo {volume}
  {107}},\ \bibinfo {pages} {186806} (\bibinfo {year} {2011})}\BibitemShut
  {NoStop}%
\bibitem [{\citenamefont {Yu}\ \emph {et~al.}(2015)\citenamefont {Yu},
  \citenamefont {Weng}, \citenamefont {Fang}, \citenamefont {Dai},\ and\
  \citenamefont {Hu}}]{Yu2015_ex}%
  \BibitemOpen
  \bibfield  {author} {\bibinfo {author} {\bibfnamefont {R.}~\bibnamefont
  {Yu}}, \bibinfo {author} {\bibfnamefont {H.}~\bibnamefont {Weng}}, \bibinfo
  {author} {\bibfnamefont {Z.}~\bibnamefont {Fang}}, \bibinfo {author}
  {\bibfnamefont {X.}~\bibnamefont {Dai}},\ and\ \bibinfo {author}
  {\bibfnamefont {X.}~\bibnamefont {Hu}},\ }\bibfield  {title} {\bibinfo
  {title} {{Topological Node-Line Semimetal and Dirac Semimetal State in
  Antiperovskite ${\mathrm{Cu}}_{3}\mathrm{PdN}$}},\ }\href
  {https://doi.org/10.1103/PhysRevLett.115.036807} {\bibfield  {journal}
  {\bibinfo  {journal} {Phys. Rev. Lett.}\ }\textbf {\bibinfo {volume} {115}},\
  \bibinfo {pages} {036807} (\bibinfo {year} {2015})}\BibitemShut {NoStop}%
\bibitem [{\citenamefont {Kim}\ \emph {et~al.}(2015)\citenamefont {Kim},
  \citenamefont {Wieder}, \citenamefont {Kane},\ and\ \citenamefont
  {Rappe}}]{Kim2015_ex}%
  \BibitemOpen
  \bibfield  {author} {\bibinfo {author} {\bibfnamefont {Y.}~\bibnamefont
  {Kim}}, \bibinfo {author} {\bibfnamefont {B.~J.}\ \bibnamefont {Wieder}},
  \bibinfo {author} {\bibfnamefont {C.~L.}\ \bibnamefont {Kane}},\ and\
  \bibinfo {author} {\bibfnamefont {A.~M.}\ \bibnamefont {Rappe}},\ }\bibfield
  {title} {\bibinfo {title} {{Dirac Line Nodes in Inversion-Symmetric
  Crystals}},\ }\href {https://doi.org/10.1103/PhysRevLett.115.036806}
  {\bibfield  {journal} {\bibinfo  {journal} {Physival Review Letters}\
  }\textbf {\bibinfo {volume} {115}},\ \bibinfo {pages} {036806} (\bibinfo
  {year} {2015})}\BibitemShut {NoStop}%
\bibitem [{\citenamefont {Chen}\ \emph {et~al.}(2015)\citenamefont {Chen},
  \citenamefont {Lu},\ and\ \citenamefont {Kee}}]{Chen2015_ex}%
  \BibitemOpen
  \bibfield  {author} {\bibinfo {author} {\bibfnamefont {Y.}~\bibnamefont
  {Chen}}, \bibinfo {author} {\bibfnamefont {Y.-M.}\ \bibnamefont {Lu}},\ and\
  \bibinfo {author} {\bibfnamefont {H.-Y.}\ \bibnamefont {Kee}},\ }\bibfield
  {title} {\bibinfo {title} {{Topological crystalline metal in orthorhombic
  perovskite iridates}},\ }\href {https://doi.org/10.1038/ncomms7593}
  {\bibfield  {journal} {\bibinfo  {journal} {Nature Communications}\ }\textbf
  {\bibinfo {volume} {6}},\ \bibinfo {pages} {6593} (\bibinfo {year}
  {2015})}\BibitemShut {NoStop}%
\bibitem [{\citenamefont {B{\"{u}}ttiker}\ \emph {et~al.}(1985)\citenamefont
  {B{\"{u}}ttiker}, \citenamefont {Imry}, \citenamefont {Landauer},\ and\
  \citenamefont {Pinhas}}]{Buttinger:1985ib}%
  \BibitemOpen
  \bibfield  {author} {\bibinfo {author} {\bibfnamefont {M.}~\bibnamefont
  {B{\"{u}}ttiker}}, \bibinfo {author} {\bibfnamefont {Y.}~\bibnamefont
  {Imry}}, \bibinfo {author} {\bibfnamefont {R.}~\bibnamefont {Landauer}},\
  and\ \bibinfo {author} {\bibfnamefont {S.}~\bibnamefont {Pinhas}},\
  }\bibfield  {title} {\bibinfo {title} {{Generalized many-channel conductance
  formula with application to small rings}},\ }\href
  {https://doi.org/10.1103/PhysRevB.31.6207} {\bibfield  {journal} {\bibinfo
  {journal} {Physical Review B}\ }\textbf {\bibinfo {volume} {31}},\ \bibinfo
  {pages} {6207} (\bibinfo {year} {1985})}\BibitemShut {NoStop}%
\bibitem [{gam()}]{gamma_conv}%
  \BibitemOpen
  \href@noop {} {}\bibinfo {note} {We use the following convention for gamma
  matrices: $\Gamma_1 = \tau_1 \sigma_1$, $\Gamma_2 = \tau_1 \sigma_2$,
  $\Gamma_3 = \tau_3$, $\Gamma_4 = \tau_1 \sigma_3$ and $\Gamma_5 = -
  \tau_2$.}\BibitemShut {Stop}%
\bibitem [{\citenamefont {Nikoli{\'{c}}}(2001)}]{Nikolic:2001cf}%
  \BibitemOpen
  \bibfield  {author} {\bibinfo {author} {\bibfnamefont {B.~K.}\ \bibnamefont
  {Nikoli{\'{c}}}},\ }\bibfield  {title} {\bibinfo {title} {{Deconstructing
  Kubo formula usage: Exact conductance of a mesoscopic system from weak to
  strong disorder limit}},\ }\href {https://doi.org/10.1103/PhysRevB.64.165303}
  {\bibfield  {journal} {\bibinfo  {journal} {Physical Review B}\ }\textbf
  {\bibinfo {volume} {64}},\ \bibinfo {pages} {165303} (\bibinfo {year}
  {2001})}\BibitemShut {NoStop}%
\bibitem [{\citenamefont {Behrends}(2018)}]{JanPhD}%
  \BibitemOpen
  \bibfield  {author} {\bibinfo {author} {\bibfnamefont {J.}~\bibnamefont
  {Behrends}},\ }\emph {\bibinfo {title} {Transport and Quantum Anomalies in
  Topological Semimetals}},\ \href
  {https://nbn-resolving.org/urn:nbn:de:bsz:14-qucosa2-331532} {Ph.D. thesis},\
  \bibinfo  {school} {Technische Universit\"at Dresden} (\bibinfo {year}
  {2018})\BibitemShut {NoStop}%
\bibitem [{Note1()}]{Note1}%
  \BibitemOpen
  \bibinfo {note} {We thank Jun-Won Rhim for this observation.}\BibitemShut
  {Stop}%
\bibitem [{\citenamefont {Noro}\ \emph {et~al.}(2010)\citenamefont {Noro},
  \citenamefont {Koshino},\ and\ \citenamefont {Ando}}]{Noro:2010ct}%
  \BibitemOpen
  \bibfield  {author} {\bibinfo {author} {\bibfnamefont {M.}~\bibnamefont
  {Noro}}, \bibinfo {author} {\bibfnamefont {M.}~\bibnamefont {Koshino}},\ and\
  \bibinfo {author} {\bibfnamefont {T.}~\bibnamefont {Ando}},\ }\bibfield
  {title} {\bibinfo {title} {{Theory of Transport in Graphene with Long-Range
  Scatterers}},\ }\href {https://doi.org/10.1143/JPSJ.79.094713} {\bibfield
  {journal} {\bibinfo  {journal} {Journal of the Physical Society of Japan}\
  }\textbf {\bibinfo {volume} {79}},\ \bibinfo {pages} {094713} (\bibinfo
  {year} {2010})}\BibitemShut {NoStop}%
\bibitem [{\citenamefont {Ominato}\ and\ \citenamefont
  {Koshino}(2014)}]{Ominato:2014ch}%
  \BibitemOpen
  \bibfield  {author} {\bibinfo {author} {\bibfnamefont {Y.}~\bibnamefont
  {Ominato}}\ and\ \bibinfo {author} {\bibfnamefont {M.}~\bibnamefont
  {Koshino}},\ }\bibfield  {title} {\bibinfo {title} {{Quantum transport in a
  three-dimensional Weyl electron system}},\ }\href
  {https://doi.org/10.1103/PhysRevB.89.054202} {\bibfield  {journal} {\bibinfo
  {journal} {Phys. Rev. B}\ }\textbf {\bibinfo {volume} {89}},\ \bibinfo
  {pages} {054202} (\bibinfo {year} {2014})}\BibitemShut {NoStop}%
\bibitem [{\citenamefont {Rammer}(2004)}]{RammerBook}%
  \BibitemOpen
  \bibfield  {author} {\bibinfo {author} {\bibfnamefont {J.}~\bibnamefont
  {Rammer}},\ }\href {https://doi.org/10.1201/9780429502835} {\emph {\bibinfo
  {title} {{Quantum Transport Theory}}}}\ (\bibinfo  {publisher} {CRC Press},\
  \bibinfo {address} {Boca Raton, FL},\ \bibinfo {year} {2004})\BibitemShut
  {NoStop}%
\bibitem [{\citenamefont {Rhim}\ and\ \citenamefont {Kim}(2016)}]{Rhim:2016dy}%
  \BibitemOpen
  \bibfield  {author} {\bibinfo {author} {\bibfnamefont {J.-W.}\ \bibnamefont
  {Rhim}}\ and\ \bibinfo {author} {\bibfnamefont {Y.~B.}\ \bibnamefont {Kim}},\
  }\bibfield  {title} {\bibinfo {title} {{Anisotropic density fluctuations,
  plasmons, and Friedel oscillations in nodal line semimetal}},\ }\href
  {https://doi.org/10.1088/1367-2630/18/4/043010} {\bibfield  {journal}
  {\bibinfo  {journal} {New Journal of Physics}\ }\textbf {\bibinfo {volume}
  {18}},\ \bibinfo {pages} {043010} (\bibinfo {year} {2016})}\BibitemShut
  {NoStop}%
\bibitem [{\citenamefont {Medvedyeva}\ \emph {et~al.}(2010)\citenamefont
  {Medvedyeva}, \citenamefont {Tworzyd{\l}o},\ and\ \citenamefont
  {Beenakker}}]{Medvedyeva:2010jf}%
  \BibitemOpen
  \bibfield  {author} {\bibinfo {author} {\bibfnamefont {M.~V.}\ \bibnamefont
  {Medvedyeva}}, \bibinfo {author} {\bibfnamefont {J.}~\bibnamefont
  {Tworzyd{\l}o}},\ and\ \bibinfo {author} {\bibfnamefont {C.~W.~J.}\
  \bibnamefont {Beenakker}},\ }\bibfield  {title} {\bibinfo {title} {{Effective
  mass and tricritical point for lattice fermions localized by a random
  mass}},\ }\href {https://doi.org/10.1103/PhysRevB.81.214203} {\bibfield
  {journal} {\bibinfo  {journal} {Physical Review B}\ }\textbf {\bibinfo
  {volume} {81}},\ \bibinfo {pages} {214203} (\bibinfo {year}
  {2010})}\BibitemShut {NoStop}%
\bibitem [{\citenamefont {Mong}\ \emph {et~al.}(2012)\citenamefont {Mong},
  \citenamefont {Bardarson},\ and\ \citenamefont {Moore}}]{Mong:2012du}%
  \BibitemOpen
  \bibfield  {author} {\bibinfo {author} {\bibfnamefont {R.~S.~K.}\
  \bibnamefont {Mong}}, \bibinfo {author} {\bibfnamefont {J.~H.}\ \bibnamefont
  {Bardarson}},\ and\ \bibinfo {author} {\bibfnamefont {J.~E.}\ \bibnamefont
  {Moore}},\ }\bibfield  {title} {\bibinfo {title} {{Quantum transport and
  two-parameter scaling at the surface of a weak topological insulator}},\
  }\href {https://doi.org/10.1103/PhysRevLett.108.076804} {\bibfield  {journal}
  {\bibinfo  {journal} {Physical Review Letters}\ }\textbf {\bibinfo {volume}
  {108}},\ \bibinfo {pages} {076804} (\bibinfo {year} {2012})}\BibitemShut
  {NoStop}%
\bibitem [{Note2()}]{Note2}%
  \BibitemOpen
  \bibinfo {note} {In class D, the system only becomes metallic for white-noise
  disorder~\cite {Medvedyeva:2010jf,Bardarson:2010jn}.}\BibitemShut {Stop}%
\bibitem [{\citenamefont {Sun}\ and\ \citenamefont
  {Syzranov}(2023)}]{Sun:2023ht}%
  \BibitemOpen
  \bibfield  {author} {\bibinfo {author} {\bibfnamefont {S.}~\bibnamefont
  {Sun}}\ and\ \bibinfo {author} {\bibfnamefont {S.}~\bibnamefont {Syzranov}},\
  }\bibfield  {title} {\bibinfo {title} {{Interactions-disorder duality and
  critical phenomena in nodal semimetals, dilute gases, and other systems}},\
  }\href {https://doi.org/10.1103/PhysRevB.108.195132} {\bibfield  {journal}
  {\bibinfo  {journal} {Physical Review B}\ }\textbf {\bibinfo {volume}
  {108}},\ \bibinfo {pages} {195132} (\bibinfo {year} {2023})}\BibitemShut
  {NoStop}%
\bibitem [{\citenamefont {Fu}\ \emph {et~al.}(2024)\citenamefont {Fu},
  \citenamefont {Guan}, \citenamefont {Xie},\ and\ \citenamefont
  {An}}]{Fu2023}%
  \BibitemOpen
  \bibfield  {author} {\bibinfo {author} {\bibfnamefont {J.-j.}\ \bibnamefont
  {Fu}}, \bibinfo {author} {\bibfnamefont {S.-t.}\ \bibnamefont {Guan}},
  \bibinfo {author} {\bibfnamefont {J.}~\bibnamefont {Xie}},\ and\ \bibinfo
  {author} {\bibfnamefont {J.}~\bibnamefont {An}},\ }\bibfield  {title}
  {\bibinfo {title} {{Quantum transport on the surfaces of topological
  nodal-line semimetals}},\ }\href {https://doi.org/10.1088/1367-2630/ad19fb}
  {\bibfield  {journal} {\bibinfo  {journal} {New Journal of Physics}\ }\textbf
  {\bibinfo {volume} {26}},\ \bibinfo {pages} {013032} (\bibinfo {year}
  {2024})}\BibitemShut {NoStop}%
\bibitem [{\citenamefont {Wang}\ and\ \citenamefont
  {Nandkishore}(2017)}]{Wang:2017be}%
  \BibitemOpen
  \bibfield  {author} {\bibinfo {author} {\bibfnamefont {Y.}~\bibnamefont
  {Wang}}\ and\ \bibinfo {author} {\bibfnamefont {R.~M.}\ \bibnamefont
  {Nandkishore}},\ }\bibfield  {title} {\bibinfo {title} {{Interplay between
  short-range correlated disorder and Coulomb interaction in nodal-line
  semimetals}},\ }\href {https://doi.org/10.1103/PhysRevB.96.115130} {\bibfield
   {journal} {\bibinfo  {journal} {Physical Review B}\ }\textbf {\bibinfo
  {volume} {96}},\ \bibinfo {pages} {115130} (\bibinfo {year}
  {2017})}\BibitemShut {NoStop}%
\bibitem [{\citenamefont {Mu{\~{n}}oz-Segovia}\ and\ \citenamefont
  {Cortijo}(2020)}]{Munoz:2020cw}%
  \BibitemOpen
  \bibfield  {author} {\bibinfo {author} {\bibfnamefont {D.}~\bibnamefont
  {Mu{\~{n}}oz-Segovia}}\ and\ \bibinfo {author} {\bibfnamefont
  {A.}~\bibnamefont {Cortijo}},\ }\bibfield  {title} {\bibinfo {title}
  {{Many-body effects in nodal-line semimetals: Correction to the optical
  conductivity}},\ }\href {https://doi.org/10.1103/PhysRevB.101.205102}
  {\bibfield  {journal} {\bibinfo  {journal} {Physical Review B}\ }\textbf
  {\bibinfo {volume} {101}},\ \bibinfo {pages} {205102} (\bibinfo {year}
  {2020})}\BibitemShut {NoStop}%
\bibitem [{\citenamefont {Kohn}(1961)}]{Kohn:1961et}%
  \BibitemOpen
  \bibfield  {author} {\bibinfo {author} {\bibfnamefont {W.}~\bibnamefont
  {Kohn}},\ }\bibfield  {title} {\bibinfo {title} {{Cyclotron Resonance and de
  Haas-van Alphen Oscillations of an Interacting Electron Gas}},\ }\href
  {https://doi.org/10.1103/PhysRev.123.1242} {\bibfield  {journal} {\bibinfo
  {journal} {Physical Review}\ }\textbf {\bibinfo {volume} {123}},\ \bibinfo
  {pages} {1242} (\bibinfo {year} {1961})}\BibitemShut {NoStop}%
\bibitem [{\citenamefont {Khmel'nitskii}(1983)}]{Khmelnitskii1983}%
  \BibitemOpen
  \bibfield  {author} {\bibinfo {author} {\bibfnamefont {D.~E.}\ \bibnamefont
  {Khmel'nitskii}},\ }\bibfield  {title} {\bibinfo {title} {{Quantization of
  Hall conductivity}},\ }\href@noop {} {\bibfield  {journal} {\bibinfo
  {journal} {Pis'ma Zh. Exp. Teor. Fiz.}\ }\textbf {\bibinfo {volume} {38}},\
  \bibinfo {pages} {454} (\bibinfo {year} {1983})},\ \bibinfo {note} {[JETP
  Lett.\ 38, 552--556 (1983)]}\BibitemShut {NoStop}%
\bibitem [{\citenamefont {Burkov}(2018)}]{Burkov:2018gc}%
  \BibitemOpen
  \bibfield  {author} {\bibinfo {author} {\bibfnamefont {A.~A.}\ \bibnamefont
  {Burkov}},\ }\bibfield  {title} {\bibinfo {title} {{Quantum anomalies in
  nodal line semimetals}},\ }\href {https://doi.org/10.1103/PhysRevB.97.165104}
  {\bibfield  {journal} {\bibinfo  {journal} {Phys. Rev. B}\ }\textbf {\bibinfo
  {volume} {97}},\ \bibinfo {pages} {165104} (\bibinfo {year}
  {2018})}\BibitemShut {NoStop}%
\bibitem [{\citenamefont {Rui}\ \emph {et~al.}(2018)\citenamefont {Rui},
  \citenamefont {Zhao},\ and\ \citenamefont {Schnyder}}]{Rui:2018jr}%
  \BibitemOpen
  \bibfield  {author} {\bibinfo {author} {\bibfnamefont {W.~B.}\ \bibnamefont
  {Rui}}, \bibinfo {author} {\bibfnamefont {Y.~X.}\ \bibnamefont {Zhao}},\ and\
  \bibinfo {author} {\bibfnamefont {A.~P.}\ \bibnamefont {Schnyder}},\
  }\bibfield  {title} {\bibinfo {title} {{Topological transport in Dirac
  nodal-line semimetals}},\ }\href {https://doi.org/10.1103/PhysRevB.97.161113}
  {\bibfield  {journal} {\bibinfo  {journal} {Physical Review B}\ }\textbf
  {\bibinfo {volume} {97}},\ \bibinfo {pages} {161113(R)} (\bibinfo {year}
  {2018})}\BibitemShut {NoStop}%
\bibitem [{\citenamefont {Bardarson}\ \emph {et~al.}(2010)\citenamefont
  {Bardarson}, \citenamefont {Medvedyeva}, \citenamefont {Tworzyd{\l}o},
  \citenamefont {Akhmerov},\ and\ \citenamefont
  {Beenakker}}]{Bardarson:2010jn}%
  \BibitemOpen
  \bibfield  {author} {\bibinfo {author} {\bibfnamefont {J.~H.}\ \bibnamefont
  {Bardarson}}, \bibinfo {author} {\bibfnamefont {M.~V.}\ \bibnamefont
  {Medvedyeva}}, \bibinfo {author} {\bibfnamefont {J.}~\bibnamefont
  {Tworzyd{\l}o}}, \bibinfo {author} {\bibfnamefont {A.~R.}\ \bibnamefont
  {Akhmerov}},\ and\ \bibinfo {author} {\bibfnamefont {C.~W.~J.}\ \bibnamefont
  {Beenakker}},\ }\bibfield  {title} {\bibinfo {title} {{Absence of a metallic
  phase in charge-neutral graphene with a random gap}},\ }\href
  {https://doi.org/10.1103/PhysRevB.81.121414} {\bibfield  {journal} {\bibinfo
  {journal} {Physical Review B}\ }\textbf {\bibinfo {volume} {81}},\ \bibinfo
  {pages} {121414(R)} (\bibinfo {year} {2010})}\BibitemShut {NoStop}%
\end{thebibliography}%

\end{document}